\documentclass[acmsmall,authorversion]{acmart}

\usepackage{float}
\usepackage{subfig}
\usepackage{listings}
\usepackage[ruled,linesnumbered]{algorithm2e}
\usepackage{xcolor}
\usepackage{tabularx}
\usepackage{multirow} 
\usepackage{multicol}

\AtBeginDocument{%
  \providecommand\BibTeX{{%
    \normalfont B\kern-0.5em{\scshape i\kern-0.25em b}\kern-0.8em\TeX}}}

\setcopyright{acmlicensed}
\copyrightyear{2026}
\acmYear{2026}
\acmJournal{TAAS}
\acmVolume{21}
\acmNumber{2}
\acmArticle{13}
\acmMonth{4}
\acmDOI{10.1145/3708998}

\newtheorem{definition}{Definition}
\begin{document}

\title{Context-Aware Proactive Self-Adaptation: A Two-Layer Model Predictive Control Approach}

\author{Zhengyin Chen}
\email{chenzy512@pku.edu.cn}
\orcid{0009-0000-8060-6851}
\affiliation{%
  \institution{Peking University}
  \streetaddress{Yiheyuan Road}
  \city{Beijing}
  \state{Beijing}
  \country{China}
  \postcode{100871}
}

\author{Jialong Li}
\affiliation{%
  \institution{Waseda University}
  \country{Japan}}
\email{lijialong@fuji.waseda.jp}

\author{Nianyu Li}
\authornote{corresponding author}
\affiliation{%
  \institution{ZGC Lab}
  \streetaddress{1}
  \city{Beijing}
  \country{China}}
\email{li_nianyu@pku.edu.cn}

\author{Wenpin Jiao}
\affiliation{%
  \institution{Peking University}
  \city{Beijing}
  \country{China}
}
\email{jwp@pku.edu.cn}

\author{Eunsuk Kang}
\affiliation{%
  \institution{Carnegie Mellon University}
  \city{Pittsburgh}
  \country{USA}
}
\email{eunsukk@andrew.cmu.edu}

\renewcommand{\shortauthors}{Chen, et al.}

\begin{abstract}
 In self-adaptive software systems, the role of context is paramount, especially for proactive self-adaptation. Current research, however, does not fully explore context's impact, for example on priorities of the requirements. To address this gap, we introduce a novel contextual goal model to capture these factors and their influence on the system. Using this, we propose a two-layer control mechanism with a context-aware model predictive control to achieve proactive adaptation for the software system and adaptation for the controller itself. By contextual prediction and a more accurate system model, our approach utilizes model predictive control to facilitate timely and efficient system adaptations, improving both performance and adaptability. Meanwhile, we perform requirement adaptation to update the contextual goal model, which in turn updates the objective function and constraints of the controller. Our experimental evaluations across two scenarios demonstrate the significant benefits of our approach in enhancing system performance. 
\end{abstract}

\ccsdesc[500]{Software and its engineering~Software design engineering}
\ccsdesc[500]{Software and its engineering~Requirements analysis}
\keywords{Proactive Adaptation, Model Predictive Control, Contextual Goal Model}

\maketitle

\section{Introduction}

In modern software systems, there is a growing demand for reliability, robustness, security, and other trustworthy properties as they need to operate autonomously in complex and dynamic environments. To address this, self-adaptation has emerged as an effective approach. Self-adaptive systems have the capability to continuously monitor the environment and the system state, enabling them to adjust their behavior and/or structure when system goals are violated \cite{lemos2013software}. However, traditional self-adaptive systems often adopt a reactive structure, whereby adaptations are made after the system goal has been violated in order to restore its satisfaction. This reactive nature may result in missed opportunities for optimal adaptation timing, leading to significant losses, particularly in critical tasks. Furthermore, frequent adaptations may be required without being able to achieve an optimal solution over time. Consequently, relying solely on reactive self-adaptation poses challenges in some scenarios.

In recent years, proactive self-adaptation has gained increasing attention among researchers in the field \cite{cooray2013proactive,moreno2015proactive,angelopoulos2016model,ayala2021prodspl,shin2021pasta}. The fundamental distinction between proactive adaptation and reactive adaptation lies in their timing of adjustment. Proactive adaptation aims to prevent potential violations by taking anticipatory actions beforehand, while reactive adaptation involves making adjustments when the goal violations are detected \cite{handte20123pc}. A proactive self-adaptive system can leverage predictive capability to anticipate future system states and proactively adapt to avert undesirable outcomes, beyond merely compensating for adverse consequences. Proactive adaptation proves particularly valuable in scenarios where the violation of system goals would result in unacceptable losses or when the effects of adaptation exhibit significant delays. 

Furthermore, control theory-based approaches have been increasingly integrated into the development of self-adaptive systems due to their solid mathematical foundation and ability to provide formal guarantees \cite{weyns2020introduction}. Among these approaches, Model Predictive Control (MPC) has emerged as a widely adopted control technique in industrial applications \cite{filieri2017control}. Utilizing a system model to forecast future behavior, MPC solves an optimization problem under specific constraints to generate optimal control inputs. This iterative process of prediction and control computation is repeated in each control loop. MPC exhibits versatility in handling multiple input and multiple output systems, aligning well with the complexities of software systems that involve multiple inputs and multiple outputs. Moreover, the proactive nature of self-adaptation, which involves predicting the future state of the system, aligns naturally with the predictive capabilities of MPC. As a result, leveraging MPC for proactive self-adaptation has been explored in several existing works \cite{angelopoulos2016model,ayala2021prodspl,wang2015qos}. 

Despite this synergy between MPC and proactive self-adaptation, existing research exhibits a notable gap: the insufficient consideration of context in MPC-based approaches. Such approaches often rely on a nominal system model that typically captures the behavior of the software system, focusing solely on the influence of the current system state and control signals on its future state, while ignoring the influence of contextual factors. Context, defined as the 
specific environmental conditions within which a system operates \cite{finkelstein2001framework}, plays a crucial role in 
shaping both the behavior and goals of self-adaptive systems, as well as the strategies employed to achieve these goals and their effectiveness of goal achievement. 
Omitting consideration of context can result in inaccurate system models, unaccounted constraints, unreasonable goals, and improper behavior, thereby diminishing the effectiveness of adaptation. 

In our previous work, we introduced a context-aware MPC approach to facilitate proactive adaptation in software systems \cite{chen2022proactive}. This approach involved incorporating contextual factors into the system model to capture their influence on the system state, performing context prediction during decision-making, and utilizing MPC to compute optimal control signals. Building upon these foundations, this paper extends the role of context in self-adaptive software systems by introducing a novel contextual goal model (CGM). 
This CGM effectively captures the dynamic context as it relates to software goals and their attainment. It utilizes control parameters to represent goal attainment, and employs performance indicators and softgoal quality functions to quantify the achievement of softgoals. Furthermore, it supports control theory-based adaptation by enabling adjustments to these control parameters. This structure allows a self-adaptive software system to flexibly adjust its goals and plan software configurations in response to dynamic environments, leveraging the capabilities of the CGM.
Leveraging this CGM, we propose a two-layer control loop consisting of an MPC Control Layer and a Meta Control Layer. The MPC Control Layer employs context-aware MPC to achieve proactive adaptation of the software system. Specifically, this layer calculates the optimal control parameters to optimize the performance of the system against varying contexts. On the other hand, the Meta Control Layer is responsible for run-time updates to the CGM in alignment with the current context. This layer also helps fine-tune the settings of the MPC, including the objective function and constraints, thus rendering the controller adaptive. By integrating contextual considerations into the control architecture, our approach enhances the software system's adaptability and performance, enabling it to dynamically adjust its behavior and goals based on the CGM and achieve more effective proactive adaptation.

The main contributions of this paper are: (1) It provides an in-depth analysis of the role that context plays in self-adaptive software systems and introduces a novel contextual goal model to encapsulate this relationship. (2) Based on the contextual goal model, we propose a two-layer control loop designed for dual adaptability: adaptation at the level of the software system and adaptive tuning of the controller itself. (3) Experiments are conducted both on the SWIM case and Smart Room cases to verify the effectiveness of the approach.

The remaining parts of this paper are organized as follows: in Section 2, we introduce some background knowledge on proactive self-adaptation, model predictive control, and goal-oriented requirement engineering as well as a brief description of a running example. In Section 3, we present the overview of our proposed two-layer framework. In Section 4, we analyze the effects of context on self-adaptive software systems and represent them using contextual goal model, followed by Section 5, where we report the design phase and run-time phase of our approach. Section 6 presents the evaluation and Section 7 elaborates on major related work. Finally, Section 8 concludes the paper along with future work.

\section{Preliminaries}
Our proposed approach synergizes Model Predictive Control (MPC) and Goal-Oriented Requirements Engineering (GORE) to facilitate proactive adaptation in software systems. In this section, we offer foundational background into each of these domains to set the stage for a deeper understanding of our research contributions. At the end of this section, we will give a running example to help the subsequent elaboration.

\subsection{Proactive Self-adaptation}
\label{sec:psa}
Classified from the temporal characteristics, self-adaptive methods can be divided into reactive and proactive \cite{salehie2009self,krupitzer2015survey}. The main difference is whether the adaptation behavior takes place before or after the adaptation need arises. Mapping the adaptation process to the MAPE loop \cite{kephart2003vision}, the basic activities for proactive adaptation also include monitoring, analyzing, planning, and executing, but with slight differences in specific behaviors. For example, in the analysis phase, for reactive adaptation, monitoring data is used to determine whether the system has accomplished its goals, while for proactive adaptation, monitoring data is used to predict the environment and the future state of the system and determine the future goal accomplishment of the system. Specifically, the decision-making process of proactive adaptation (analysis and planning phase) incorporates the key principles of (1) Using historical data to predict future changes in the environment (2) Predicting the attainment of the system's future goals based on the current state of the system and predictions about the future of the environment (3) After that, these predictions guide the planning of adaptation actions, which aims to avoid the goal violations or improve the performance in the future.

Although proactive adaptation requires additional overhead, more complex adaptation logic, and the potential risk of malicious adaptation, it offers distinct advantages over reactive adaptation in certain scenarios. By anticipating and addressing potential goal violations or performance degradation before they occur, software systems have the opportunity to prevent or minimize the impact of such violations. This is particularly beneficial in safety-critical systems, real-time applications, and mission-critical environments, where users prioritize minimizing violations, minimizing losses, and minimizing recovery time after violations.

\subsection{Model Predictive Control}
\label{sec:mpc}
\begin{figure}[!t]
  \centering
  \includegraphics[width=0.8\linewidth]{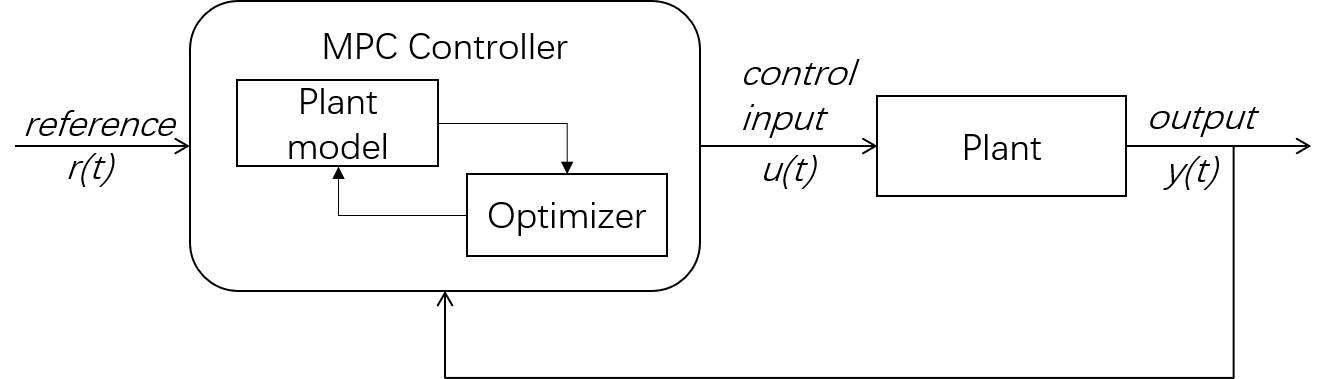}
  \caption{Overview of Model Predictive Control}
  \label{fig:mpc}

\end{figure}

Model Predictive Control is a widely used modern control strategy as it offers a compromise between optimality and computation cost \cite{kouvaritakis2016model}. The overview process of MPC is shown in Figure \ref{fig:mpc}. It uses a system dynamic model (e.g. Equation \eqref{equ:sysmodel}) to represent the relationship of the system state $x(t+1)$ to the previous system state $x(t)$ and control signals $u(t)$, as well as the relationship of the system output $y(t)$ to the system state $x(t)$ and control signals $u(t)$ (where $t$ indicates the time).

\begin{equation}
\label{equ:sysmodel}
\begin{aligned}
    x(t+1) &= f(x(t),u(t)) \\
    y(t) &= g(x(t),u(t))
\end{aligned}
\end{equation}

In each control loop, the controller is required to solve a constrained optimization problem, shown in Equation \eqref{equ:opt}, that is to find a sequence of control signals $u$ that make the objective function $J$ optimal over a horizon $H$ under certain constraints. This objective function usually aims to bring the output of the system $y$ as close as possible to a reference value $r$, while considering the cost of the control signals $u$. The predicted system outputs need to be calculated through the system dynamic model. In Equation \eqref{equ:opt}, the objective function is the quadratic form, which is the most common form in MPC, and the $P,Q$ denote the weights of each term. If the system model is of linear form, then this optimization problem is a quadratic programming problem, with many existing solvers.
\begin{equation}
\label{equ:opt}
    \begin{aligned}
        minimize_{u(i)}\ &\ J=\sum_{i=0}^{H-1} Q(y(i)-r(i))^2 +  Pu(i)^2 \\
        subject\ to\ &\ x(i+1) = f(x(i),u(i)) \\
                     &\ y(i) = g(x(i),u(i)) \\
                     &\ u_{min}\le u(i) \le u_{max},\\
                     &\ y_{min}\le y(i) \le y_{max},\\
                     &\ i=0,1,2,...,H-1 ,\\
                     &\ x(0)=x(t)
    \end{aligned}   
\end{equation}

In general, MPC follows the receding horizon principle, that is, in each control loop, the controller will calculate a sequence of control signals, but the controller will only retain the first term of this sequence as a control signal and discard the rest, and repeat the calculation process in the subsequent control loop using the newly measured system state. Therefore, MPC is also known as Receding Horizon Control. Due to its ability to solve optimization problems with constraints that can help improve the long-term performance of systems, MPC has already been used to implement proactive self-adaptive systems.

\subsection{Goal-Oriented Requirement Engineering}
\label{sec:gore}
\begin{figure}[!t]
  \centering
  \includegraphics[width=0.9\linewidth]{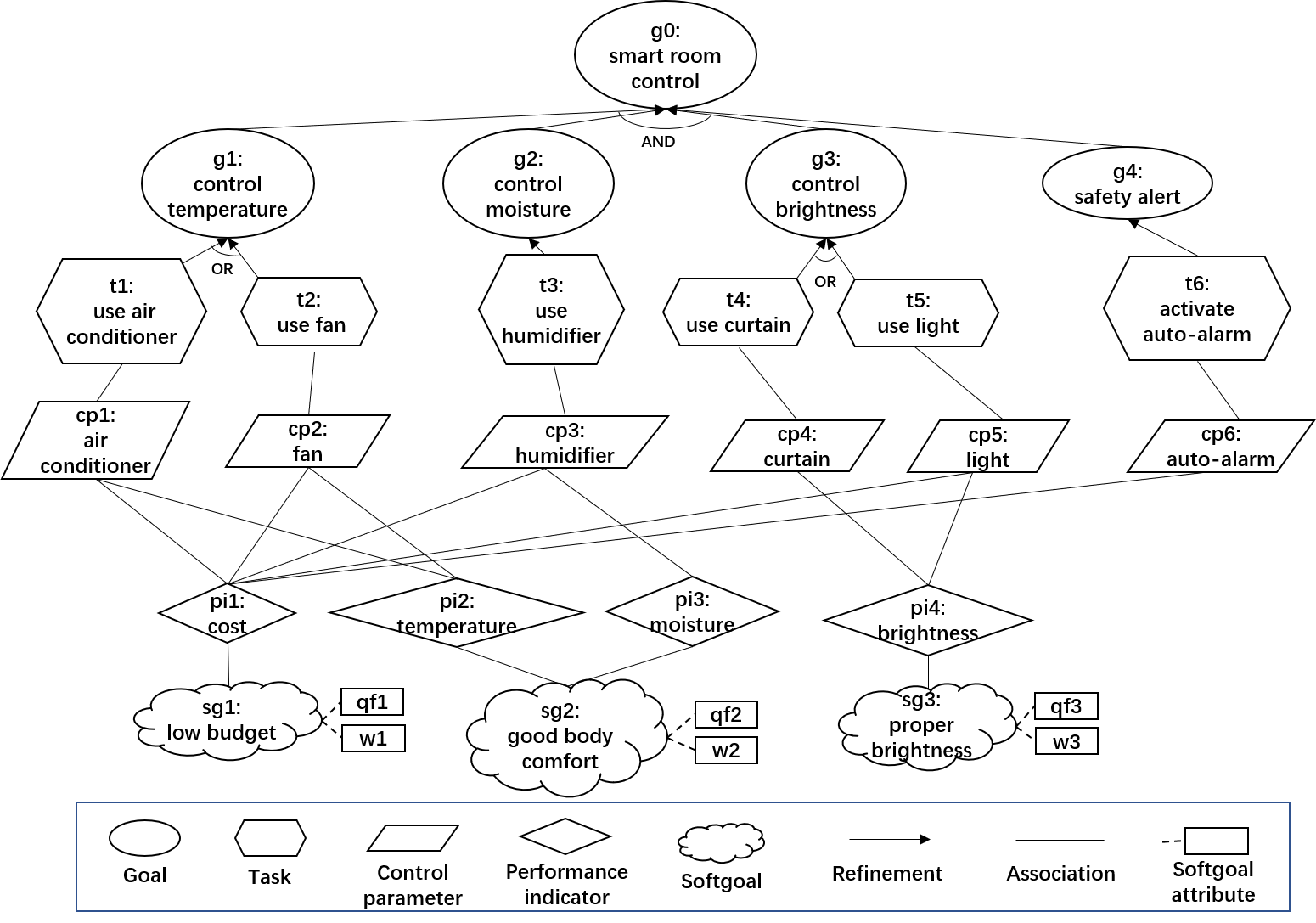}
  \caption{Goal Model of A Smart Room Control System}
  \label{fig:gm}

\end{figure}
Requirements engineering is the initial phase of the software development life cycle, encompassing the definition of domain logic, identification of stakeholders' requirements, and the documentation of relevant information for subsequent analysis and implementation \cite{nuseibeh2000requirements}. In recent years, Goal-Oriented Requirements Engineering (GORE) has become the mainstream method in requirements engineering due to its ability to provide a precise and complete requirements specification, improve the readability of requirements, and facilitate the exploration of alternative system proposals \cite{van2001goal,ali2010goal}. GORE employs a goal model to refine goals and subsequently refine them into tasks that can be assigned to agents, while softgoals are used to represent the non-functional requirements of the software system. 

The goal model concepts are based on previous works \cite{silva2012requirements}, including: 
\begin{itemize}
    \item Goals: Functional requirements that the software system needs to complete. 
    \item Tasks: Actions that can be directly executed to achieve some goals.
    \item AND/OR Refinement: Refinement is the connection between two elements. AND/OR has obvious semantics, an AND-refinement means that in order to satisfy the parent goal, all of its children must be satisfied, while for an OR-refinement, only one child needs to be attained. Specifically, we use decomposition to denote the link between goals and goals, and operationalization to denote the link between goals and tasks.
    \item Softgoals: Softgoal represents some non-functional requirements of the software system. Unlike ordinary goals, softgoals are rarely considered to be fully completed, but as close as possible to the user's satisfaction. Conflicts may arise between different softgoals, so weights must be assigned to each softgoal to represent its priority. Each softgoal is associated with a quality function to describe the satisfaction level of the softgoal. The quality function specifies the relationship between the performance indicators related to a softgoal and the level of satisfaction.
    \item Performance Indicators: Performance indicators are measurable system states in the software system, used to indicate the performance and status of the software system. The satisfaction of a softgoal can be related to one or more performance indicators. 
    \item Control Parameters: Control parameters are variables that can be modified to alter the behavior or configuration of the system. They are often operated by tasks and can be adjusted to achieve specific requirements. The value of each performance indicator is affected by one or more control parameters.
\end{itemize}

\subsection{Running Example}
\label{sec:run-exp}
In this article, we use a Smart Room Control System as a running example. Figure \ref{fig:gm} presents the goal model of this system, aiming to enable autonomous control of intelligent devices within a room on behalf of the user. The system comprises four subgoals, namely \emph{temperature control, moisture control, brightness control, and safety alert}, which can be achieved through various tasks that adjust different control parameters. For example, the goal \emph{temperature control} can be attained by utilizing task \emph{use air-conditioner}, by adjusting the control parameter \emph{air conditioner} to attain the desired temperature.

The system encompasses four core performance indicators: \emph{cost, temperature, moisture, and brightness}, with values accessible via sensors or other monitoring methods. These indicators are influenced by control parameters, and their relationship are defined by the system model (i.e. the system dynamic model in MPC).

The system also includes three softgoals, \emph{low budget, good body comfort, and proper brightness}. These softgoals aim to enhance user experience while ensuring optimal device control.
Each softgoal, tied to specific indicators, has a \emph{quality function} (qf) and a \emph{weight} (w). For instance, \emph{low budget} is linked to \emph{cost}, with its qf measuring user satisfaction inversely related to cost. A higher weight to \emph{low budget} prompts the system to favor cost-effective control parameters during adaptation.

At present, our goal model does not account for contextual factors and their potential impacts. In self-adaptive software systems, like a Smart Room Control System, contextual influence is significant. It not only affects the performance indicators, but also affects the goals, tasks,  softgoal priorities, and softgoal quality functions. For instance, while we manipulate the temperature using control parameters like \emph{air conditioner or fan}, contextual factors like outside temperature substantially influence the outcome. Additionally, specific contexts, such as extreme weather conditions or particular user activities, could lead to an increased priority for the softgoal \emph{good body comfort}. 

To address this, we propose an enhancement of the goal model to incorporate contextual factors, transitioning to a contextual goal model. This ensures that the MPC controller is context-responsive, enabling it to self-update in varied contexts, enhancing its efficacy in goal attainment.

\section{Approach Overview}

\begin{figure}[!t]
  \centering
  \includegraphics[width=\linewidth]{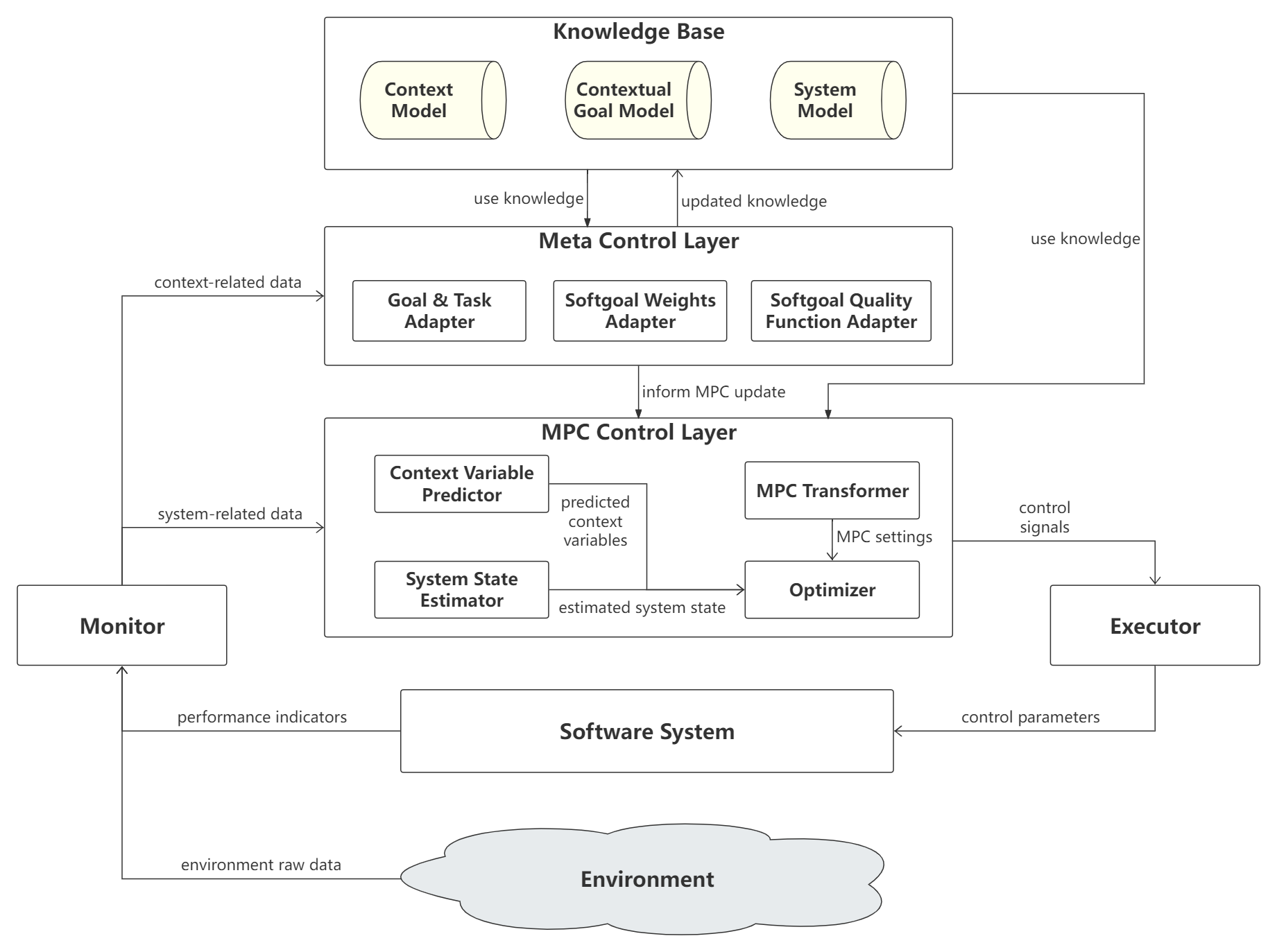}
  \caption{Two Layer Control Framework Overview}
  \label{fig:framework}
\end{figure}

To address the insufficient consideration of the context in existing MPC-based approaches, we propose a two-layer control framework based on MAPE-K model. The illustrative overview of the framework is depicted in Figure \ref{fig:framework}. Central to our approach is the incorporation of a CGM, which serves to map the influence of context on the software system. Our framework further refines the analysis and planning stages of MAPE-K into two specialized layers: the Meta Control Layer and the MPC Control Layer. 
In this design, the MPC Control Layer concentrates on managing the effects of contextual shifts on the system's performance indicators. Utilizing context-aware MPC, this layer calculates optimal control parameters for proactive self-adaptation. Conversely, the Meta Control Layer addresses the influence of context on the system requirements by continually updating the CGM based on the current context. It also instructs the MPC Control Layer to modify its settings in alignment with these updates.

During the design phase, three key models within the knowledge base are constructed. The \textbf{Context Model} outlines the specific context variables and context specification. The \textbf{Contextual Goal Model} defines the goals that must be met to satisfy user requirements and elucidates how these goals are influenced by the context. The \textbf{System Model} details the relationships between the system outputs, states, control inputs, and context, forming the basis for predicting future system states. Following the establishment of these models, the \textbf{MPC Transformer} in the MPC Control Layer initializes the optimizer, taking into account the default context and the CGM.

At runtime, the \textbf{Monitor} is tasked with overseeing both the environmental conditions and the internal state of the software system. It collects raw environmental data and key performance indicators from the system. After data preprocessing, it directs two sets of information to different layers: system-related data, encompassing performance indicators and their associated context variables, is passed to the MPC Control Layer, while context-specific data is routed to the Meta Control Layer.

The Meta Control Layer is responsible for adjusting the CGM to the evolving contextual conditions and guiding the MPC Control Layer to update its settings. This layer is subdivided into three functional components: \textbf{the goal/task adapter, the softgoal weights adapter, and the softgoal quality function adapter}. Each of these is responsible for adjusting specific elements of the CGM. Once adjustments are made, the updated CGM is stored in the Knowledge Base, and an instruction is sent to the MPC Control Layer to reconfigure the MPC settings accordingly. Additionally, the context-related data collected by the monitor is integrated into the knowledge base in sync with updates to the CGM.

The MPC Control Layer, on the other hand, is responsible for proactive adaptation to contextual changes, thereby ensuring continuous alignment with user requirements. Within this layer, the \textbf{MPC transformer} configures the optimizer according to the CGM, while the \textbf{Context Variable Predictor}, the \textbf{System State Estimator} and the \textbf{Optimizer} collaborate to execute the proactive adaptation decision-making process. This decision-making process adheres to the principles in Section \ref{sec:psa}: (1) The \textbf{Context Variable Predictor} forecasts the future values of the context variables utilizing the \textbf{Context Model}. These predicted values, along with the most recent system state provided by the \textbf{System State Estimator}, serve as critical inputs for the optimizer. (2) Within the \textbf{Optimizer}, future goal attainment will be predicted leveraging the \textbf{System Model} and the input. (3) Based on this prediction, the \textbf{Optimizer} solves for optimal control parameters to guarantee the goal achievement during the control horizon. Then it passes these control signals to the \textbf{Executor}.

The \textbf{Executor} transforms control signals generated by the MPC Control Layer into actionable control parameters for the software system. Subsequently, it instructs the software system to carry out the necessary adaptations in alignment with these parameters.

Our research is specifically designed to investigate how dynamic context affects both the performance indicators and the requirements of a software system. Given this targeted focus, we have intentionally excluded analyses related to the system's impact on the external environment, as such considerations fall outside the scope of our current research. Furthermore, it's worth noting the differentiated use of predictive capabilities across the framework's layers. At the MPC Control Layer, we leverage the predictive elements of both the context model and the system model to enable proactive adaptation of the software system's control parameters. In contrast, the Meta Control Layer does not employ context prediction. Instead, it retains the CGM in static throughout a control cycle for two primary reasons. First, the constraints of the existing MPC solver prevent any alteration of the objective function during a single solution cycle. Second, we currently lack a compensatory mechanism, similar to MPC's receding horizon principle, to offset prediction errors that could potentially lead to requirement inaccuracies. 

\section{Analyzing Contextual Influence Using Goal Model}
\label{sec:CGM}
In this section, we delve into the ways in which context influences self-adaptive software systems. Section \ref{sec:cm} initiates the definition of "context" as employed within the scope of this article. Following this, Section \ref{sec:ctx-inf} scrutinizes the contextual influence on various constituents of the goal model, building upon the goal model elaborated in Section \ref{sec:gore}. Finally, Section \ref{sec:cgm} integrates these contextual considerations into the existing goal model, thereby culminating in the formulation of a comprehensive contextual goal model.

\subsection{Context Model}
\label{sec:cm}

The term "context" is employed across various domains and can have divergent meanings. In the field of software engineering, it is commonly defined as "A context is a partial state of the world that is relevant to an actor’s goals" \cite{ali2010goal}. 
Following this definition and combining it with the notion of "environment" in self-adaptive systems \cite{weyns2020introduction}, we further refine the concept of context to be "a partial state of the external world in which the system operates and interacts, and which is relevant to the system’s goals."  This encompasses elements such as the physical environment, computational resources, user behavior, etc.

To formally specify context, we first define a set of context variables, which are objects in the external world directly observable by the software system. These variables encapsulate contextual information pertinent to the system. Within the MAPE-K model, the Monitor component retrieves the values of these variables using sensors or other methods.
A context variable $cv$ should have the following attributes:
\begin{itemize}
    \item \textbf{Name}: A descriptor of the context variable.
    \item \textbf{Type}: Indicates the data type of the variable, such as numerical, boolean, or enumeration.
    \item \textbf{Value}: Denoted by $cv(t)$ where $t$ represents time, reflecting the variable's value at a specific moment within its range. 
    \item \textbf{Prediction Function}:Essential for proactive adaptation, this function is defined as  $Pred: cv(t-l),...,cv(t-1)\rightarrow cv(t+1),...,cv(t+h)$, where $cv(t-l),...,cv(t-1)$ represents the historical values of the context variable spanning a length of $l$, while $cv(t+1),...,cv(t+h)$ denotes the predicted future values of the context variable extending over a period of $h$. In this paper, we assume that each context variable is independent, which means the future value of a context variable is only related to the historical value of the context variable itself.
\end{itemize}

Further, we adopt the specification of context from Ali et al. \cite{ali2010goal}, which posits that context is a formula composed of world predicates. These world predicates are predicates about context variables, which describe specific properties of partial states of the world. The syntax for this formula is provided in Equation \eqref{equ:formula}. 
The context variables we use in the running example are shown in Table \ref{tbl:cv} and the context specification is shown in Table \ref{tbl:c}. For example, the context "weather is extreme" can be expressed as "outside temperatures greater than 30" or "outside temperatures less than 10". In this example, the context variable is the \emph{outside temperature}, and the world predicates are "outside temperature greater than 30" and "outside temperature less than 10."

\begin{table}[!t]
\renewcommand{\arraystretch}{1}
\caption{Context variables in Smart Room Control System}
\label{tbl:cv}
\centering
\begin{tabular}{cccc}
\toprule
context variable & name & type & Value Range \\
\midrule
cv1 & Connectivity & enumeration& \{off, on\} \\

cv2 & Outside brightness & numerical & [0,100] \\

cv3 & User Activity & enumeration & \{idle, work, sleep, away\} \\

cv4 & Outside Temperature & numerical & [-30,50] \\

cv5 & Outside Moisture & numerical & [0,1] \\
\bottomrule
\end{tabular}
\end{table}

\begin{table}[!t]
\renewcommand{\arraystretch}{1}
\caption{Context Specification in Smart Room Control System}
\label{tbl:c}
\centering
\begin{tabular}{cccc}
\toprule
context & related context variable & formula \\
\midrule
c1(network is available) & cv1(connectivity) & cv1 = on \\

c2(user is sleeping) & cv3(user activity) & cv3 = sleep\\

c3(user is working) & cv3(user activity) & cv3 = work \\

c4(user is away) & cv3(user activity) & cv3 = away \\

c5(weather is extreme) & cv5(outside temperature) & cv > 30 or cv5 < 10 \\
\bottomrule
\end{tabular}
\end{table}

\begin{equation}
\label{equ:formula}
    Formula := \text{World}\ \text{Predicate} \mid (Formula) \mid Formula\ AND\ Formula \mid Formula\ OR\ Formula
\end{equation}

The process of analyzing the monitored environment and defining contexts in terms of variables and their values is called \emph{context operationalization}, which is conducted by the Meta Control Layer in our framework. However, the details of the process of context variables definition and context operationalization are beyond the scope of this paper. More information could be acquired in literature \cite{knauss2016acon,rodrigues2019enhancing}. 

During the execution of the software system, the values of context variables can change within their specified value ranges, and such changes are often uncontrollable, which leads to a dynamic and uncertain environment that affects the system's goals. To manage this dynamicity and uncertainty at runtime and ensure that the software system continues to meet its users' requirements, it is necessary to incorporate the influence of context in the goal model to aid runtime adaptation.

\subsection{Analysis of the Influence of Context}
\label{sec:ctx-inf}

\begin{figure*}[!t]
\centering
\subfloat[Goal]{\includegraphics[width=0.2\textwidth]{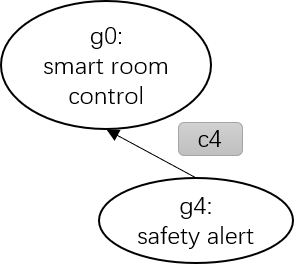}
\label{fig_inf1}}
\hfil
\subfloat[Task]{\includegraphics[width=0.2\textwidth]{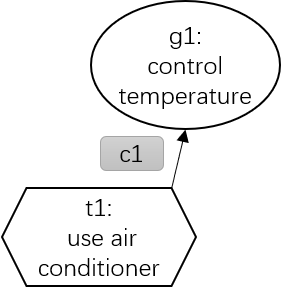}
\label{fig_inf2}} 
\hfil
\subfloat[Softgoal Weight]{\includegraphics[width=0.3\textwidth]{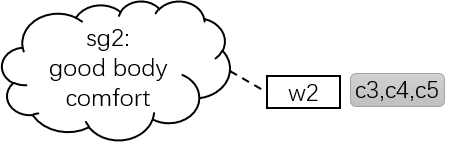}
\label{fig_inf3}} \\
\hfil
\subfloat[Softgoal Quality Function]{\includegraphics[width=0.3\textwidth]{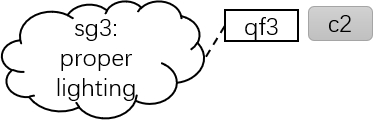}
\label{fig_inf4}} 
\hfil
\subfloat[Performance Indicator]{\includegraphics[width=0.35\textwidth]{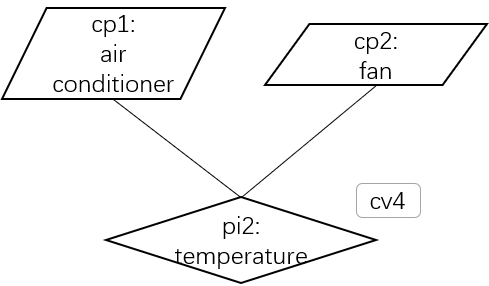}
\label{fig_inf5}} 
\caption{Influence of context on requirements}
\label{fig_inf}
\end{figure*}

Building on the goal model outlined in the previous section, we can systematically analyze the influence of context on the requirements based on the affected objects, which are:

\begin{itemize}
    \item Goal (and Refinement): Context can be the precondition for a goal, which means that a goal is only active in a specific context. In AND refinement, when the context holds, a goal must be achieved to achieve its parent goal; in OR refinement, when the context holds, realizing its parent goal can be achieved by this goal. For example, goal g4 (safety alert) activates when context c4 (user is not home) holds, as depicted in Figure \ref{fig_inf1}. As g4 is obtained by an AND refinement of g0, and if c4 does not hold, then the realization of g0 no longer needs g4 to be realized.
    
    \item Context influences task in a similar manner as goal, with AND/OR scenarios. 
    In AND scenarios, achieving a goal requires executing a task if a specific context holds, while in OR scenarios, a goal can be achieved by executing a task if the specific context holds. For instance, task t1 (use air conditioner) requires context c1 (network is available) holds, as depicted in Figure \ref{fig_inf2}. Since t1 is obtained by an OR refinement of g1, if c1 does not hold, then g1 cannot be realized through t1, but only through t2.
   
    \item Softgoals weight: In varying contexts, users have distinct preferences, leading to changing priorities among softgoals. For instance, in a scenario with three softgoals: \emph{low budget}, \emph{good body comfort}, and \emph{proper brightness}, the user might prioritize good body comfort and proper brightness while working, i.e. c3 holds, as depicted in Figure \ref{fig_inf3}. 
    
    \item Softgoals quality function: In various contexts, the user's satisfaction with softgoals achievement is assessed differently. For instance, with the softgoal sg3 (proper brightness), suitable brightness criterion depends on the \emph{user activity}. When sleeping (c2 holds), satisfaction is negatively correlated with brightness intensity, but when working (c3 holds), satisfaction is positively correlated, as depicted in Figure \ref{fig_inf4}.

    \item Performance indicator: Context, in conjunction with control parameters, impacts the software system's performance indicators. As the context varies, the performance indicators can alter even if control parameters remain consistent. For easier integration in MPC, this effect is represented using context variables. This showcases the quantitative relationship between performance indicators, control parameters, and context variables. For instance, room temperature is influenced by both the control parameters, such as the air conditioner and fan, and the context variable—\emph{outside temperature}, as shown in Figure \ref{fig_inf5}.

\end{itemize}

\subsection{Contextual Goal Model}
\label{sec:cgm}

\begin{figure*}[!t]
\centering
\includegraphics[width=0.9\textwidth]{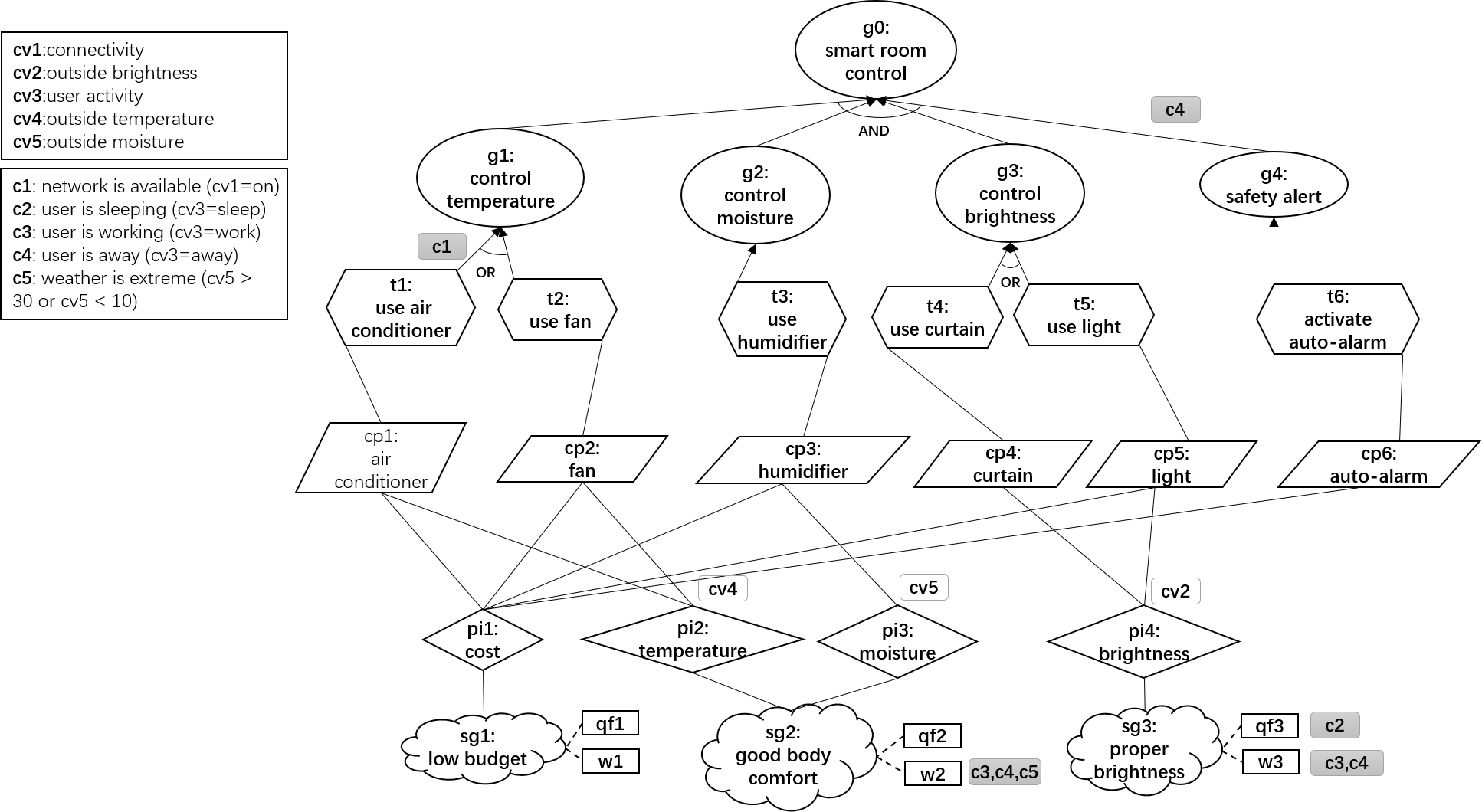}
\caption{Contextual Goal Model of Smart Room Control System}
\label{fig:cgm}
\end{figure*}

Following the analysis above, we present the CGM of the Smart Room Control System in Figure \ref{fig:cgm}. Furthermore, the CGM of a self-adaptive software system is formally defined as: 

\begin{definition}
$CGM=(CV,C,G,T,PI,SG,CP,R,A,AC)$
\begin{itemize}
    \item $CV=\{cv_1,cv_2,...,cv_{n_{cv}}\}$ represents the context variables associated with the software system.
    \item $C=\{c_1,c_2,...,c_{n_{c}}\}$ represents the context after context operationalization with the context variables.
    \item $G=\{g_1,g_2,...,g_{n_{g}}\}$ represents all the goals of the software system.
    \item $T=\{t_1,t_2,...,t_{n_{t}}\}$ represents all the tasks of the software system.
     \item $PI=\{pi_1,pi_2,...,pi_{n_{pi}}\}$ represents all the performance indicators in the software system.
    \item $SG=\{sg_1,sg_2,...,sg_{n_{sg}}\}$ represents all the softgoals of the software system. Each $sg_i$ has two attributes, $(w_i,qf_i)$, $w_i$ is the weight for $sg_i$, indicates its priority, $w_i$ is influenced by $C_{sg_i},C_{sg_i}=\{c_j\mid (sg_i,c_j)\in AC_{sg}\}$; $qf_i$ is the quality function for $sg_i$, indicates the satisfaction level of softgoal for the given values of the performance indicators.
    \item $CP=\{cp_1,cp_2,...,cp_{n_{cp}}\}$ represents all the control parameters of the software system, each control parameter $cp_i$ has a current value and a value range $[cp_{i,min},cp_{i,max}]$.
    \item $R=(R_d,R_o)$ represents the Refinement relation. Where $R_d=\{(g_i,g_j,AND\vert OR,c_k)\mid g_i,g_j\in G,(g_j,c_k)\in AC_g\}$ represents the decomposition between goals and goals with possible contextual influence. $R_o=\{(g_i,t_j,AND\vert OR,c_k)\mid g_i\in G,t_j\in T,(t_j,c_k)\in AC_t\}$ represents the operationalization relationship between goals and tasks with possible contextual influence. 
    \item $A=(A_t,A_{pi},A_{sg})$ represents the association relationship of elements in the goal model. 
    \begin{itemize}
        \item $A_t=\{(t_i,cp_j)\mid t_i\in T,cp_j\in CP\}$ indicates what control parameter a task is associated with, i.e. what control parameters the task changes.
        \item $A_{pi}=\{(pi_i,cp_j)\mid pi_i\in PI,cp_j\in CP\}$ indicates what control parameters are associated with a performance indicator, i.e. how the control parameters affect the value of the performance indicator. 
        \item $A_{sg}=\{(sg_i,pi_j)\mid sg_i\in SG,pi_j\in PI\}$ indicates what performance indicators a softgoal is associated with, i.e. what performance indicators reflect the realization of the softgoal.
    \end{itemize}
    \item $AC=\{AC_g,AC_t,AC_{pi},AC_{sg}\}$ represents how the contexts are associated with other elements in the CGM. 
    \begin{itemize}
        \item $AC_g=\{(g_i,c_j)\mid g_i\in G, c_j\in C\}$, indicates which contexts the goal is associated with. 
        \item $AC_t=\{(t_i,c_j)\mid t_i\in T, c_j\in C\}$,  indicates which contexts the task is associated with.
        
        \item $AC_{sg}=\{(sg_i,c_j)\mid sg_i\in SG, c_j\in C\}$, indicates which contexts the softgoal is associated with. Specifically, the context influences the weight $w_{sg_i}$ of the softgoal $sg_i$. The assessment of the softgoal satisfaction level is also influenced by the context, denoted as $qf_{sg_i}=f(PI_{sg_i},C_{sg_i}), PI_{sg_i}=\{pi_j\mid (sg_i,pi_j)\in A_{sg}\},C_{sg_i}=\{c_j\mid (sg_i,c_j)\in AC_{sg}\}$.
        \item $AC_{pi}=\{(pi_i,cv_j)\mid pi_i\in PI, cv_j\in CV\}$, indicates which context variables the performance indicator is associated with. Specifically, the calculations of performance indicators are influenced by context variables, represented as $pi_i=f(CP_{pi_i},CV_{pi_i}), CP_{pi_i}=\{cp_j\mid (pi_i,cp_j)\in A_{pi}\}, CV_{pi_i}=\{cv_j\mid (pi_i,cv_j)\in AC_{pi}\}$.
    \end{itemize}
\end{itemize}
\end{definition}

Compared with the existing CGMs in self-adaptive systems, our CGM has the following advantages: (1) We represent the relation between dynamic context and goal (and softgoal) attainment in the CGM to support control theory-based adaptation. We introduce control parameters to represent goal attainment and introduce performance indicators and softgoal quality functions to quantify soft goal attainment in the CGM. This enables the seamless application of control theory-based adaptation by adjusting control parameters to ensure goal attainment and optimize softgoal satisfaction. (2) We comprehensively analyze the context influences on the software system, including influences on goals, tasks, softgoals, and performance indicators. By incorporating contextual conditions on goals and tasks, and user preferences on softgoals into the CGM, the software system can flexibly adjust the CGM to different contexts at runtime. This enhances the realization of user requirements.

Given a graphical representation of the CGM, we can convert it to a formal representation, making it more convenient for implementing the self-adaptive systems. During the design phase, the CGM can help in the design of the MPC controller. At runtime, the Meta Control Layer can represent changes in requirements by updating the CGM, while the MPC Control Layer updates the MPC settings with the new CGM. More details will be presented in the next section.

\section{Context-aware Two Layer Adaptation}
In this section, we are more specific about the details of the two control layers, including how the two control layers are designed based on the CGM, and how the two control layers respond to the influence of contextual changes on the software system at runtime. We will use the Smart Room Control System to help explain. A brief description of this case is in Section \ref{sec:run-exp}, while the CGM is provided in Section \ref{sec:cgm}. We assume that the CGM has been acquired before designing the controller.

\subsection{Design based on Contextual Goal Model}
\label{sec:design}
\subsubsection{MPC Control Layer design}
\label{sec:mpc-design}
In \cite{filieri2015software}, the authors discuss the development of a control loop for a software system, which includes seven main steps: (1) Identify the goals. (2) Identify the knobs. (3) Devise the model. (4) Design the controller. (5) Formally evaluating the closed loop system. (6) Implement and integrate the controller. (7) Test and validate the system. Here we mainly focus on the first four steps, since the subsequent steps are about the implementation and the verification and are not the focus of this paper. We do not introduce new methods for validation, instead, experimental verification in two cases is provided in Section \ref{sec:exp}. Specifically, we delve into the design of an MPC controller (the basics of MPC are introduced in Section \ref{sec:mpc}) in accordance with this process, while also exploring the role of the CGM introduced in the preceding section to facilitate the design procedure.

\paragraph{Identify the Goals}

The first step in the process involves establishing clear and quantifiable goals that the software system aims to achieve. Typically, the goal of the controller is expressed as the setpoint of the system output, i.e. the reference values of performance indicators in CGM. In the CGM, we use goals to represent functional requirements, while softgoals are used to capture non-functional requirements. A self-adaptive software system needs to fulfill all the goals while also maximizing the satisfaction of softgoals. Therefore, we transform the goals into the constraints on the control parameters and convert the softgoals to minimize the error between the performance indicators and their optimal values.

A goal is achieved by completing the tasks obtained from the goal refinement, and the values of the corresponding control parameters determine the completion of these tasks. %
For AND refinement, achieving the parent goal necessitates that all child goals' constraints are met. While for OR refinement, satisfying just one child goal's constraints is adequate. 
The satisfaction of softgoals is closely related to specific performance indicators. We can determine the optimal values for these performance indicators, the values that maximize the quality function, through the defined quality function. Hence, given a CGM for a software system, the optimal value $r_{pi_j}$ of the performance indicators $pi_j$ should be $r_{pi_j}=argmax_{pi_j}\{qf_i\}, (sg_i,pi_j)\in A_{sg}$. 

For example, to achieve the goal of \emph{control temperature}, we utilize \emph{OR} refinement connecting to control parameters \emph{air conditioner} and \emph{fan}. No additional constraints are imposed; ensuring at least one parameter value within its valid range is sufficient. The same applies to the goal \emph{control brightness}. For \emph{control moisture}, it's streamlined, involving only one control parameter. The \emph{safety alert} goal is realized by setting the \emph{auto-alarm} control parameter to \emph{on}. 
To optimally meet the softgoal of \emph{low budget, good body comfort, and proper brightness}, we determine the values of performance indicators that maximize the corresponding quality function. The reference values for these indicators are set at $r_{pi_1}=0,r_{pi_2}=25,r_{pi_3}=0.5,r_{pi_4}=100$.

\paragraph{Identify the Knobs}
The second step involves identifying the adjustable knobs that can be manipulated to take action and fulfill the goals. In the CGM, the control parameters represent the components or aspects of the software system that can be modified, and these parameters can be directly associated with the knobs. It is worth noting that control parameters in the software system may exist in various forms, and the control signal generated by the controller needs to be a quantifiable value. This necessitates a coding transformation or mapping process to convert the control parameters into a suitable representation for the control signal. 

In the Smart Room example, six control parameters are identified. $cp_1$(air conditioner), $cp_2$(fan), and $cp_3$(humidifier) are numeric and directly correlate with adjustable knobs. In contrast, $cp_4$(curtain), $cp_5$(light), $cp_6$(auto-alarm), with values of $\{\text{on},\text{off}\}$, require conversion, for instance, to $\{0,1\}$.

\paragraph{Devise the Model}

The next step in the process entails developing a suitable model for the software system under control, which captures the relationship between changes in knob values and the desired goals, i.e. the impact of control parameters on performance indicators. In MPC, the system model is typically formulated as a state space model (Equation \eqref{equ:sysmodel}). This model represents the dynamic behavior of the software system by describing how the system outputs evolve over time in response to control inputs. The most common linear form of the state space model is shown in Equation (\ref{equ-ssmodel}):

\begin{equation}
\label{equ-ssmodel}
\begin{aligned}
     {\mathbf{x}}(t+1)&=A\mathbf{x}(t)+B\mathbf{u}(t) \\
    \mathbf{y}(t)&=C\mathbf{x}(t)+D\mathbf{u}(t)
\end{aligned}
\end{equation}

where $t$ denotes time, $\mathbf{y}$ denotes the system outputs, $\mathbf{x}$ denotes the system states, and $\mathbf{u}$ denotes the control signals. $A, B, C, D$ are the coefficient matrices.

According to the CGM in Section \ref{sec:CGM}, for the general case, the system outputs $\mathbf{y}$ are the performance indicators of the software system, i.e. $\mathbf{y}=[pi_1,\dots,pi_{n_{pi}}]$. The control signals $\mathbf{u}$ are the control parameters that affect the performance indicators, i.e. $\mathbf{u}=[cp_1,\dots,cp_{n_{cp}}]$. The system states $\mathbf{x}$ are not directly mapping to the element in CGM, it could be just the abstract variables that connect inputs and outputs. In this case, the $D$ matrix is usually the zero matrix, considering that the system outputs are directly determined by the system states and are not directly related to the control signals. The dimension of $A,B,C$ should be $n_x*n_x,n_x*n_u,n_{pi}*n_x$ accordingly. In some scenarios, the behavior of the software system is not necessarily linear, and at this point, through the system states $\mathbf{x}$, it is also possible to approximate the complex nonlinear relationship through a linear form.

For the simplest system structure, the system states $\mathbf{x}$ are the same as system outputs $\mathbf{y}$, which means the dynamics of the system can be seen as the performance indicators and control parameters that affect the performance indicators. At this time, the second equation in the state space model can be simplified to $\mathbf{y}(t)=\mathbf{x}(t)$ and only need to consider ${\mathbf{x}}(t+1)=A\mathbf{x}(t)+B\mathbf{u}(t)$ one equation.

However, this form does not represent the influence of context on performance indicators, so we use a modified form of the state space model, as shown in Equation \eqref{equ-mssmodel}.

\begin{equation}
\label{equ-mssmodel}
\begin{aligned}
     {\mathbf{x}}(t+1)&=A\mathbf{x}(t)+B\mathbf{u}(t)+E\mathbf{cv}(t) \\
    \mathbf{y}(t)&=C\mathbf{x}(t)+D\mathbf{u}(t)
\end{aligned}
\end{equation}

where $\mathbf{cv}$ represents the context variables, $E$ is the coefficient matrix corresponding to the context variable. The context is added to the system model as a separate item to reflect the relations between the system states and the context variables. %

This form is inspired by the system model form in Robust Model Predictive Control \cite{bemporad1999robust}, which adds the disturbance as a separate item to the system dynamic model. The context is similar to the disturbance, but the context has its own characteristics. First of all, although the system cannot directly control the context, it can obtain the values of the context variables, that is to say, the context is measurable to the system. Secondly, although the change of the context is random, it also has certain laws, and we can predict the change of the context. Therefore, this allows us to directly incorporate contextual factors into the system model. On the other hand, the disturbance term is usually related to measurement error or noise. It needs to be considered independently for different software systems, while the context affecting the software system is objective for all software systems. %

Regarding how to build a system model, system identification \cite{ljung1998system} is a widely used method for building a black-box model. System identification is to input specific signals to the system and collect the corresponding output results, and then estimate the coefficient matrix in the system model according to the collected input and output data. When we identify the software system, we need to obtain $<\mathbf{u}, \mathbf{y}>$ values as training data through simulation or other methods. The estimation process is to first determine the order of the model, that is, the number of system states $\mathbf{x}$. Generally, we select the smallest order that provides the most key information through Hankel singular values. Then, it is necessary to find a set of optimal coefficient matrices to minimize the error between the system output predicted by the model and the actual system output. The possible search methods include the Gauss-Newton method, the Levenberg-Marquardt method, and the gradient descent method.

When using the context-aware system dynamic model, the above-mentioned system identification method can also be used to estimate the system model, but two more steps are required: (1) In addition to the control input (control parameters) and system output (performance indicators), we also need to obtain the value of the context variables as an item of training data, the training data becomes in the form of $<\mathbf{u, cv, y}>$ at this time; (2) We need to convert the model as Equation \eqref{equ:ss-tran} so that it can be converted into the model form required by the general system identification method. This transformation uses the idea of the block matrix as the dimensions of the matrices are correct. It is noted that the model of this form is only used in the stage of system identification for convenient calculation. To make the model more accurate, we can set the coefficients to be fixed at 0 if there is no relationship between the corresponding terms. That is, given a CGM, $b_{i,j}$ in coefficient matrix $\mathbf{B}$ is not fixed at 0 only if $(pi_i,cp_j) \in A_{pi}$, and $e_{i,j}$ in coefficient matrix $\mathbf{E}$ is not fixed at 0 only if $(pi_i,cv_j) \in AC_{pi}$.

\begin{equation}
\label{equ:ss-tran}
\begin{aligned}
    {\mathbf{x}}(t+1)&=A\mathbf{x}(t)+B\mathbf{u}(t)+E\mathbf{cv}(t) \\
    &=A\mathbf{x}(t)+\begin{bmatrix}B&E\end{bmatrix}\begin{bmatrix}\mathbf{u}(t) \\ \mathbf{cv}(t)\end{bmatrix}\\
    &=A\mathbf{x}(t)+B\mathbf{'u'}(t)
\end{aligned}
\end{equation}

For example, the performance indicator \emph{temperature}, is influenced by the previous moment's temperature, two control parameters, and an external context variable - the outside temperature. The state space equation for this setup is articulated as:  $pi_2(t+1)=a_{2,2}*pi_2(t)+b_{2,1}*cp_1(t)+b_{2,2}*cp_2(t)+e_{2,4}*cv_4(t)$.

\paragraph{Design the Controller}

In the design of an MPC controller, several aspects need to be considered, including defining the objective function, setting constraints, and configuring other relevant parameters associated with the controller. Here we focus on objective function and constraints.

In MPC, the objective function to be optimized typically involves quantifying the error between the predicted output and a predefined goal. To comprehensively capture the trade-offs between cost and benefit, the control input may also be included in the objective function. In general, the objective function can be represented as Equation (\ref{equ-objfunc}):
\begin{equation}
\label{equ-objfunc}
    J=\sum_{i=1}^{H}[\mathbf{y}(t+i)-\mathbf{r}(t+i)]^{T}Q[\mathbf{y}(t+i)-\mathbf{r}(t+i)]+{\mathbf{u}(t+i)}^{T}P\mathbf{u}(t+i)
\end{equation}
where $\mathbf{y}(t+i)$ is the predicted performance indicators at time $t+i$, obtained from the system model, $\mathbf{r}(t+i)$ is the reference value for performance indicators. $Q$ and $P$ are the matrices used to control the weights. The superscript $T$ denotes the transpose of the matrix. $H$ is the time horizon for prediction and control. This form is similar to Equation \eqref{equ:opt}, except that here we use a matrix representation to reflect multiple performance indicators and multiple control parameters.
As mentioned in the \emph{Identify the Goals} step, by analyzing the softgoal's quality function we can get the optimal value for the performance indicators and use this value as the reference value for the output, while the weight matrix $Q, P$ can be obtained from the softgoals weights. 

When considering constraints, it is necessary to account for two types of constraints. Firstly, constraints need to be imposed on the valid range of control parameter values. These constraints ensure that the control inputs remain within acceptable bounds and adhere to system-specific limitations, such as the upper bound for the control parameter \emph{air conditioner}. Secondly, constraints arising from the achievement of goals in the CGM must also be taken into consideration which we have discussed in the \emph{Identify the Goals} step in this subsection. 

After designing the objective function and constraints, the optimization problem (e.g. Equation \eqref{equ:opt}) that the controller needs to solve is defined. In each control loop, the controller obtains the latest system state, then it can use the system model to solve this optimization problem and obtain the optimal sequence of control parameters.

For example, in the Smart Room case, we used the performance indicator \emph{cost} to represent the overhead of control parameters, so we used the four performance indicators in CGM directly in the objective function and omitted the term ${\mathbf{u}(t+i)}^{T}P\mathbf{u}(t+i)$. The acquisition of their reference values is described in the \emph{Identify the Goals} step. Suppose the weights of the three softgoals in our case are, $w_1=0.2,w_2=0.4,w_3=0.4$, and the corresponding performance indicators have weights $q_1=0.2,q_2=0.2,q_3=0.2,q_4=0.4$, since $pi_2,pi_3$ jointly determine the $sg_2$. Thus $Q$ is a diagonal matrix with values on the diagonal of 0.2, 0.2, 0.2, 0.4.

\subsubsection{Meta Control Layer design}
\label{sec:meta-design}
In the design phase, softgoal adaptation rules need to be designed for the Meta Control Layer. Since the Meta Control Layer needs to adapt the softgoal's priority and the softgoal's quality function according to the varying contexts, in this paper, we use the rule-based adaptation approach, and the adaptation rules can be of the following forms $<If\ Context\ c_i , Then\ use\ weights [w_1,.... .w_q]>$ and $<If\ Context\ c_i, Then\ use\ qf_j\ for\ sg_k>$. 

Regarding rule design, it requires user involvement since both softgoal prioritization and quality functions are strongly correlated with the user's own preferences. Some techniques will help to automatically generate rules with little user participation, for example, AHP method \cite{serral2017contextual} and learning method \cite{li2023preference}\cite{lunarejo2021requirements}. In this paper, we do not focus on the rule generation techniques, we assume that these rules are generated by domain experts.

For example, a weights adaptation rule and a quality function adaptation rule could be: 

\[
<If\ c2(user\ is\ sleeping), Then\ use\ weights\ [0.5,0.3,0.2]> 
\]
\[
<If\ c2(user\ is\ sleeping), Then\ use\ \{qf:s=1-pi_4/ threshold\}\ for\ sg_3>
\]

\subsection{Two-Layer Runtime Control}
In describing the runtime adaptation process for the two control layers, we use a context change scenario within the Smart Room Control System to illustrate how a Smart Room system equipped with two adaptation layers will adapt. The adaptation scenario is: the outside temperature is steadily increasing, with no substantial variations in outside moisture or brightness. Additionally, the user has transitioned from working to sleeping, and an unexpected interruption has disrupted the network connection. We first introduce the adaptation process of the Meta Control Layer, and then the MPC Control Layer, which is in accordance with the normal adaptation order of a self-adaptive software system.

\subsubsection{Meta Control Layer}

\begin{algorithm}
\footnotesize
\caption{Meta-layer adaptation}
\label{alg:req-adapt}
\KwIn{C: Current context, CGM: The original CGM}
\KwOut{CGM': Updated CGM}
{\color{olive}\tcp{goal,task adaptation}}
\ForEach{$d_r\ in \ CGM.D_r$}{
    \If{d.c is not valid}{
        $CGM\gets CGM.filter(d.g_j)$;
    }
}
\ForEach{$d_o\ in \ CGM.D_o$}{
    \If{d.c is not valid}{
        $CGM\gets CGM.filter(d.t_j)$;
    }
}

{\color{olive}\tcp{softgoal weights adaptation}}
$validWeights\gets []$;\\
\ForEach{$rule\ in\ weightAdaptRules$}{
    \If{rule.c is valid}{
        $validWeights.append(rule.w)$;
    }    
}
$weights\gets means(validWeights)$;\\
$CGM.setWeights(weights)$;\\
{\color{olive}\tcp{softgoal quality function adaptation}}
\ForEach{$rule\ in\ qfAdaptRules$}{
    \If{rule.c is valid}{
        $CGM.setQF(rule.qf)$;
    }
}
\Return CGM
\end{algorithm}

\begin{figure*}[!t]
\centering
\includegraphics[width=0.9\textwidth]{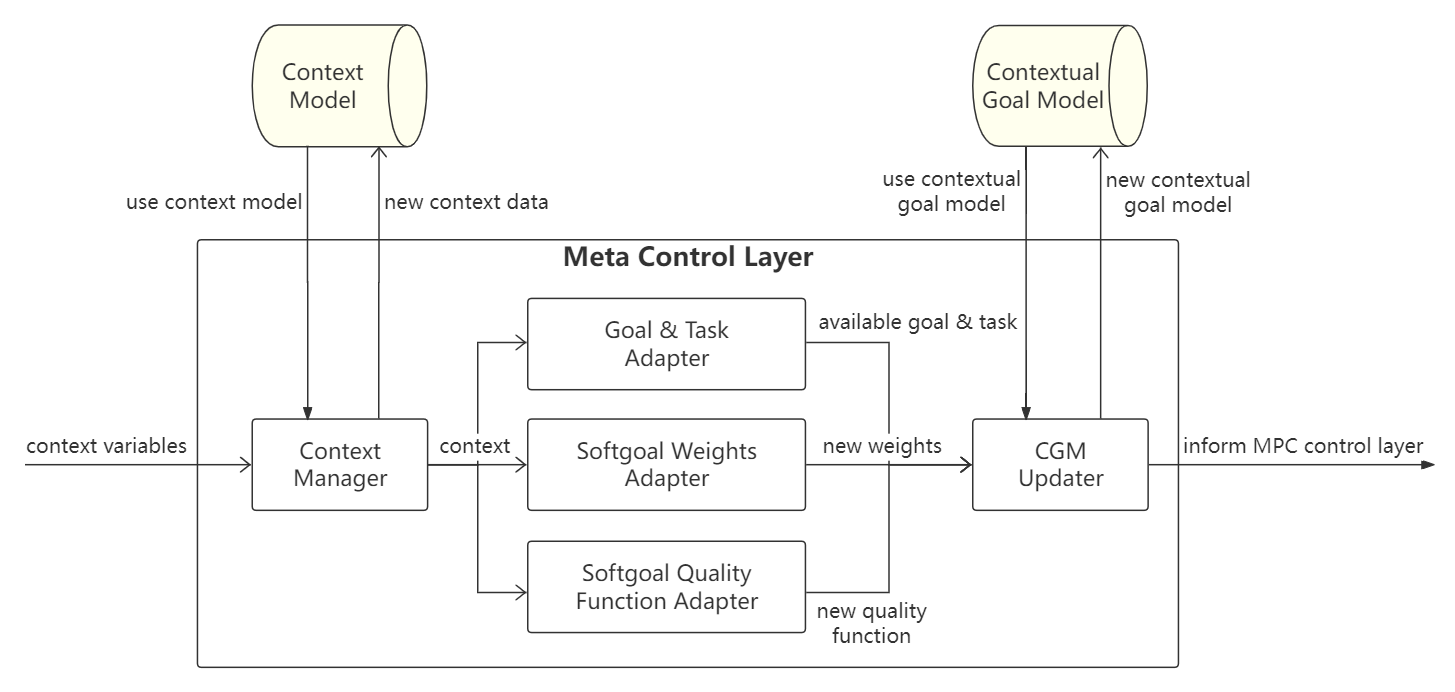}
\caption{Meta Control Layer}
\label{fig:metalayer}
\end{figure*}

The Meta Control Layer is responsible for adjusting the CGM so that the controller can adapt to the influence of context changes on goals, tasks, and softgoals and guide the MPC Control Layer to select the most appropriate settings. It includes a Context Manager, three Adapters, and a CGM updater, as depicted in Figure \ref{fig:metalayer}. 

The control process in the Meta Control Layer is as follows: 
\begin{description}
    \item[(1)] Firstly, the Meta Control Layer needs to perform context operationalization based on the context model and context variable data delivered by the Monitor to obtain the latest context. Since there is no mechanism similar to the receding horizon principle at the Meta Control Layer to mitigate the impact of misprediction and misadaptation, at this layer we do not perform context prediction but rather adapt to the current context. After the context operationalization, the original contexts [c1, c3] are changed into a new context [c2].
    
    \item[(2)] The adaptation in the Meta Control Layer is shown in Algorithm \ref{alg:req-adapt}. In line 1-10, the Goal \& Task Adapter performs goal and task filtering, those that are not available in the current context are removed to reduce the space of the CGM. The process of selecting appropriate goals and tasks is as described in \cite{ali2010goal}. In this case, given that context c1 no longer applies, task t1 becomes unavailable and is consequently removed from the CGM. In line 11-16, the Softgoal Weights Adapter selects the currently applicable set of weights and averages them to obtain the final weights according to the current context and weights adaptation rules. In line 19-23, the Softgoal Quality Function Adapter selects the applicable quality function according to the current context and quality function adaptation rules. In the updated context where c2 holds but c3 does not, the corresponding softgoal weights adaptation rule and softgoal quality function adaptation rule are activated, which are given in Section \ref{sec:meta-design}. The weights of the three softgoals are adjusted to [0.5, 0.3, 0.2] and the quality function of the softgoal \emph{proper brightness} is modified to $qf:s=1-pi_4/ threshold$.
    
    \item[(3)] After completing the above three processes, a new CGM will be generated, and the Meta Control Layer will update this CGM to the knowledge base and notify the MPC Control Layer to perform the update of the MPC settings.
\end{description}

\subsubsection{MPC Control Layer}

\begin{algorithm}
\caption{MPC Transformer}\label{alg:mpctrans}
\footnotesize
\KwIn{CGM: Contextual Goal Model for software system}
\KwOut{mpc: An initialized MPC controller}

{\color{olive}\tcp{Parameters needed to update the MPC planner}}
$lower\_bounds\gets []$;\\
$upper\_bounds\gets []$;\\
$references \gets []$; \\
$weights \gets []$; \\

{\color{olive}\tcp{Recursively get control parameters constraints through goals refinement}}
$lower\_bounds, upper\_bounds\gets GetConstraints(CGM.g_0)$

{\color{olive}\tcp{Obtain the reference value of the system output through the softgoal quality function, and obtain the weights of the system output through the softgoal weights}}
\ForEach{$sg_i\ in\ CGM.SG$}{
$pi\_list=findAllPerformanceIndicator(sg_i)$;\\
$ref\_list=argmax_{pi_j\ in\ pi\_list}(sg_i.qf)$; \\
\ForEach{$r_j$\ in\ $ref\_list$}{
$references.append(r_j)$; \\
}
\ForEach{$pi_j\ in\ pi\_list$}{
    $weights.append(sg_i.w/|pi\_list|)$
}
}

$mpc \gets MPCInit(references,weights,constraints)$; \\
\Return $mpc$;\\
\end{algorithm}

\begin{algorithm}
\caption{GetConstraints}\label{alg:refine}
\footnotesize
\KwIn{g: A goal in CGM}  
\KwOut{lower\_bounds: lower bounds of the control parameters to achieve goal g, upper\_bounds: upper bounds of the control parameters to achieve goal g} 

$lower\_bounds\gets [cp_{1,MIN},...,cp_{n_{cp},MIN}]$;\\
$upper\_bounds\gets [cp_{1,MAX},...,cp_{n_{cp},MAX}]$;\\
$subgoals = findAllSubGoals(g)$;\\
\If{$subgoals\ is\ empty$}{ {\color{olive}\tcp{no child goals, find the constraints on control parameters through tasks}}
    $tasks = findAllTasks(g)$;\\
    \If{$refinetype\ is\ AND$}{ {\color{olive}\tcp{For AND refinement, take the intersection of the upper and lower bounds of the control parameters corresponding to all tasks}}
        \ForEach{$t_i\ in\ tasks$}{
            $lower\_bounds\gets max(lower\_bounds,t_i.lower\_bounds)$;\\
            $upper\_bounds\gets min(upper\_bounds,t_i.upper\_bounds)$;\\
        }
    }
    \Else{
        {\color{olive}\tcp{For OR refinement, the upper and lower bounds of the control parameters corresponding to all tasks are taken as the union set}}     
        \ForEach{$t_i\ in\ tasks$}{
            $lower\_bounds\gets min(lower\_bounds,t_i.lower\_bounds)$;\\
            $upper\_bounds\gets max(upper\_bounds,t_i.upper\_bounds)$;\\
        }
    }
    \Return $lower\_bounds,upper\_bounds$;
}
\Else{ {\color{olive}\tcp{has child goals, obtain constraints from child goals}}
    \ForEach{$g_i\ in\ subgoals$}{
        $child\_lower\_bounds\gets GetConstraints(g_i)$;\\
        \If{$refinetype\ is\ AND$}{ 
            $lower\_bounds\gets max(lower\_bounds,child\_lower\_bounds)$;\\
            $upper\_bounds\gets min(upper\_bounds,child\_upper\_bounds)$;\\
        }
        \Else{ 
            $lower\_bounds\gets min(lower\_bounds,child\_lower\_bounds)$;\\
            $upper\_bounds\gets max(upper\_bounds,child\_upper\_bounds)$;\\
        }
    }
\Return $lower\_bounds,upper\_bounds$;
} 

\end{algorithm}

\begin{figure*}[!t]
\centering
\includegraphics[width=0.9\textwidth]{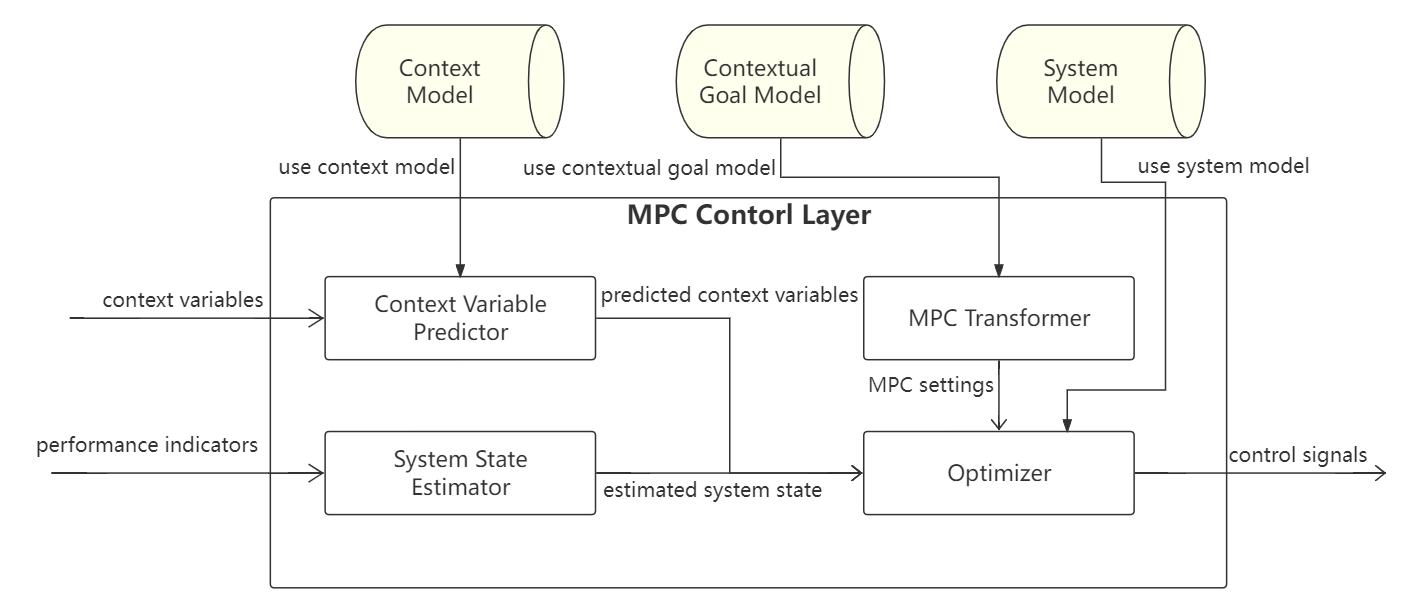}
\caption{MPC Control Layer}
\label{fig:mpclayer}
\end{figure*}

The MPC Control Layer is responsible for proactively adapting to the influence of contextual changes on performance indicators at runtime, selecting the optimal control parameters to guarantee that the software system achieves all its goals while maximizing the satisfaction of softgoals. It mainly includes a System State Estimator, a Context Variable Predictor, an MPC transformer, and an Optimizer, as depicted in Figure \ref{fig:mpclayer}.

The control process in MPC Control Layer is as follows: 
\begin{description}
    \item[(1)] When the system starts to run, the MPC transformer initializes the settings of the optimizer based on the CGM or updates the settings if the CGM has been updated. In our example, the Meta Control Layer has updated the CGM thus the MPC transformer needs to update the MPC settings. This process is similar to the design process in Section \ref{sec:mpc-design}, as shown in Algorithm \ref{alg:mpctrans}. The variables needed to initialize an MPC are defined in lines 1-4. In line 5, the constraints needed to achieve the goals are obtained step by step through the refinement of the root goal, this function \emph{GetConstraints} is defined in Algorithm \ref{alg:refine} in detail. In this function lines 4-19 deal with goals that can be refined directly into tasks, choosing different ways of generating constraints through AND/OR refinement, while lines 20-33 recursively generate their own constraints from the constraints of the subgoals, again AND/OR in different ways. The value of the control parameter \emph{air conditioner} will be restricted to 0 due to task t1 being unavailable. Back to Algorithm \ref{alg:mpctrans}, in lines 6-15, the performance indicators' reference values and corresponding weights in the MPC objective function are obtained through the softgoals' attribute in the CGM. The reference value of the performance indicator \emph{brightness} in the objective function will be changed from 100 to 0, and the weights of the performance indicators will be updated to [0.5,0.15,0.15,0.2] according to the new CGM. Finally, the initialization of the mpc is performed in line 16;
    \item[(2.1)] The System State Estimator collects the performance indicators ($y$) of the software system and converts them into system state ($x$) to pass to the optimizer as input. 
    Typically, the Kalman filter approach \cite{kalman1960anew} can be used to convert measured performance indicators into estimated system states.
    In the example case, the System State Estimator directly conveys the performance indicators to the Optimizer. 
    \item[(2.2)] The Context Variable Predictor obtains the context data from the Monitor and uses the context model in the Knowledge Base to make context predictions and passes the predicted values to the Optimizer as input. In this case, we employ time-series prediction methods, such as ARIMA, to predict future values of context variables, including \emph{outside temperature, outside moisture, and outside brightness} as they are associated with the performance indicators. These predicted values are then forwarded to the Optimizer.
    \item[(3)] The Optimizer solves an optimization problem with constraints based on the input latest system state and context prediction values, using the system model, to obtain the optimal sequence of control parameters. In this paper, this optimization problem is a quadratic programming problem (QP problem) with constraints, for this problem, some solvers already exist.  For example, our MPC implementation uses the CasADi tool to solve the optimization problem, where the default solver is Ipopt (Interior Point Optimizer). While in the widely used Matlab's MPC toolbox, three QP solvers are provided, namely Active-set solver, Interior-point solver, and ADMM solver.
    \item[(4)] According to the receding horizon principle, only the first item is applied to the software system, and the subsequent items are discarded. In this instance, the "air conditioner" is set to 0, the "fan" is adjusted to 2, and the "light" is turned off in accordance with the updated control parameters.
\end{description}

The receding horizon principle is used because the prediction of the future system state at the current moment is not accurate, and the controller will solve the optimization problem again according to the latest obtained system state at the next moment to obtain a new solution. It can mitigate the impact of incorrect predictions to some extent. In real applications, the controller may have a limited solution time, and the previously obtained solution can be saved for use when the solution cannot be obtained within the limited time or cannot find available solutions.

\section{Experiments}
\label{sec:exp}
Our experiments are centered around three main research questions:

RQ1: \emph{Effectiveness of the MPC Control Layer}. What is the effectiveness of the context-aware MPC in the MPC Control Layer? Does it improve the effectiveness compared to the base MPC?

RQ2: \emph{Effectiveness of the Meta Control Layer}. Does the requirement adaptation in the Meta Control Layer improve the effectiveness as opposed to using solely the MPC Control Layer?

RQ3: \emph{Efficiency of the two-layer approach}. What time overhead do the two control layers introduce?

\subsection{Experiment Setup}
We have conducted our experiments utilizing simulations from the Smart Room exemplar\footnote{https://github.com/easton-chen/SmartHome-exp.git} and the SWIM case\footnote{https://github.com/easton-chen/SWIM-2Layer.git} \cite{moreno2018swim}. The Smart Room exemplar is implemented in Python, while the SWIM case comes from an existing open-source setup. We've made slight modifications to the latter to align it with our methodology. %

For each case, we predefined the context variables in the environment and constructed the CGM based on domain knowledge. For context variables that may affect performance indicators, we use the ARIMA model for prediction, which is a popular time series prediction model, and we use Python's statsmodels library for implementation. Subsequently, we designed the MPC controller and the adaptation rules for the Meta Control Layer as described in Section \ref{sec:design}. For the system identification process, we used Matlab's System Identification Toolbox\footnote{https://it.mathworks.com/products/sysid.html}, and for the MPC implementation, we used Python's do-mpc package version 4.0.0  \cite{fiedler2023dompc}, which supports time-varying parameters in the system model (i.e., the context variables) and utilizes CasADi \cite{Andersson2019} and IPOPT\footnote{https://coin-or.github.io/Ipopt/index.html} as the optimization problem solver.

To assess the effectiveness of the controller, we focus on how well the software system achieves its goals and softgoals as defined in the CGM. In our experiment cases, achieving the goal primarily requires that control parameters remain within a legal range, so our focus is more on optimizing the softgoals. The effectiveness is quantified by evaluating the performance indicators. Specifically, we examine the SASO properties of the controller, which are standard metrics in control theory-based adaptation methods \cite{HellersteinDPT04}. Additionally, we use the satisfaction level of softgoals as an overall metric for evaluating system performance. The definitions of these metrics are:

\begin{itemize}
    \item \emph{Stability} is a binary property that tells if the system output converges to a constant value. 
    \item \emph{Accuracy} is related to the performance indicators, it is defined as the average error with respect to the reference. Therefore, a smaller value indicates better accuracy, which also applies to settling time and overshoot.
    \item \emph{Settling time} is the minimum time for converging to a stable solution.
    \item \emph{Overshoot} also related to the performance indicators, it is defined as the maximum error during the transient.
    \item \emph{Satisfaction level} is the weighted sum of softgoals' quality function values, denoted as $S=\sum_{i=1}^{n_{sg}} w_{sg_i}*qf_{sg_i}$. Thus a higher value indicates better satisfaction.
\end{itemize}

To begin with, we examined the SASO properties and goal fulfillment of the entire system. For comparison, we used the base MPC method, similar to the approaches in \cite{angelopoulos2018engineering, ayala2021prodspl}. This method lacks contextual insights, meaning it does not incorporate context prediction, excludes contextual factors from the system model, and is unable to adapt goals in response to context shifts.

Coming up next, we evaluated the effectiveness by conducting experiments for individual control layers in accordance with RQ1 and RQ2.
In addressing RQ1, we also compare our approach to the base MPC approach.  
The main difference is that our MPC Control Layer incorporates a context prediction model and a context-aware system model to more accurately estimate the system performance.
For RQ2, we juxtaposed the performance of a singular control layer (solely the MPC Control Layer) against that of a dual control layer setup (encompassing both the MPC Control Layer and the Meta Control Layer). The main difference is that the Meta Control Layer can address the influence of context on user requirements, subsequently recalibrating the software system's goals. 
For RQ3, we collected the time needed for our controller to compute optimal control signals (t1) and the duration taken to update the MPC settings (t2). 

The specific settings of each group of experiments is presented separately in each case.

\subsection{Smart Room}
\subsubsection{Experiment setting}
\begin{figure*}[!t]
\centering
\subfloat[Low Budget]{\includegraphics[width=0.3\textwidth]{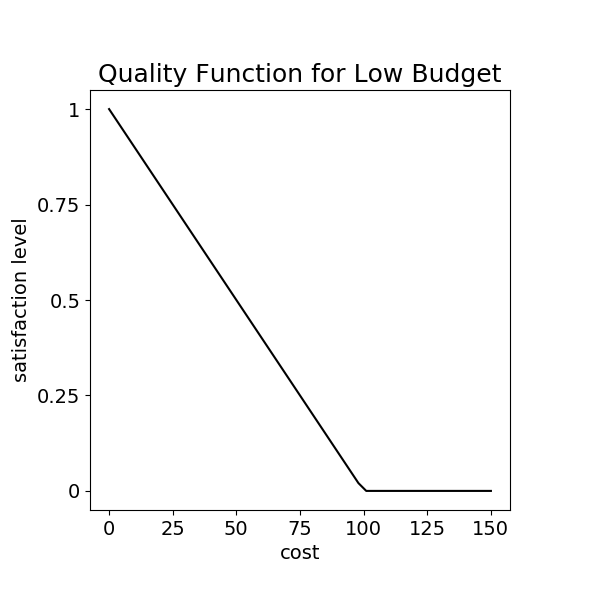}
\label{fig_srqf1}}
\hfil
\subfloat[Proper brightness]{\includegraphics[width=0.3\textwidth]{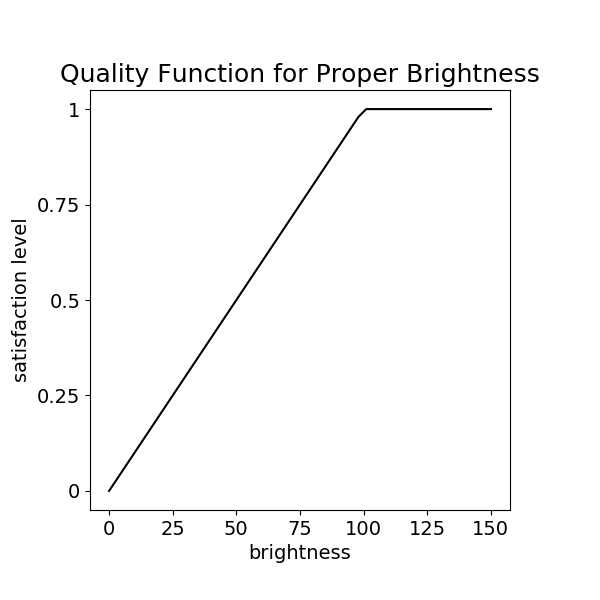}
\label{fig_srqf2}} 
\hfil
\subfloat[Good Body Comfort]{\includegraphics[width=0.3\textwidth]{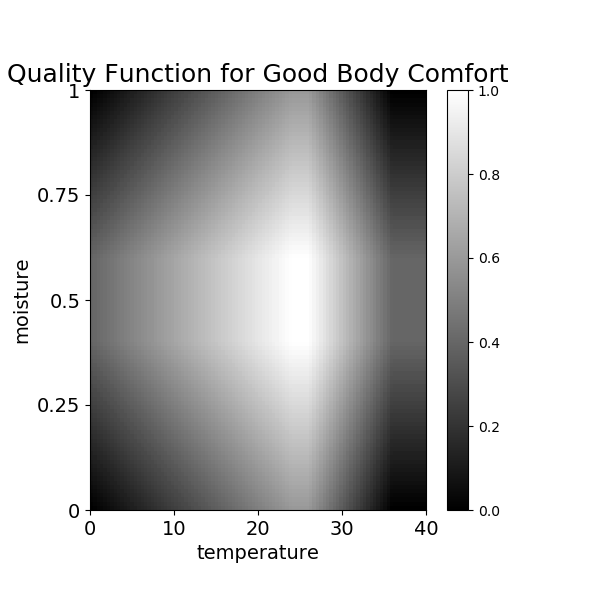}
\label{fig_srqf3}} 
\caption{Quality Functions for Softgoals}
\label{fig_srqf}
\end{figure*}

The Smart Room Control System is introduced in Section \ref{sec:run-exp}, with its CGM detailed in Section \ref{sec:cgm}. This system is designed to autonomously manage devices in a room on behalf of the user.  It seeks to achieve goals related to temperature control, moisture control, brightness adjustments, and safety alerts. In doing so, it must also satisfy three softgoals: maintaining a low budget, ensuring good body comfort, and providing appropriate brightness levels. The default quality function for each softgoal is depicted in Figure \ref{fig_srqf}, which indicates how performance indicators correlate with the satisfaction of individual softgoals. Performance indicators like temperature, moisture, and brightness levels are directly sourced from sensors, while the performance indicator cost is determined based on the present control parameters. Each softgoal is assigned an equal default weight. The system's overall satisfaction level is determined as previously discussed, while both weights and the quality function may vary in different contexts.

During our experiments, context variable values are derived in the following way: for the context variables \emph{outside temperature} and \emph{outside moisture}, we first produce a random mean and standard deviation. These values are then modulated in a trigonometric pattern. For instance, outside temperatures might peak during the day and dip at night; for the context variable \emph{connection}, it is generated entirely at random; for \emph{outside brightness}, it is designed to exhibit random fluctuations during daylight hours but remains constant at night; and for \emph{user activity}, it is also designed to changes randomly throughout the day and stabilizes at a constant value during nighttime.

As for the experiments of SASO properties, given the dynamic nature of the environment in which the self-adaptive systems operate, the system may need to continually adjust in response to evolving contexts, thus calculating SASO metrics directly may not be feasible. Therefore, we design a specific experiment where context alterations occur at a distinct point in time, allowing us to test the system's adaptive capabilities both at initial setup and upon context shifts. In this set of experiments, we chose three scenarios. Scenario 1 corresponds to the adaptation from the initial state to stabilization, where all control parameters are initially set to 0 or off, the user behavior is idle, the outside moisture is 0.5, the outside temperature is 30 degrees Celsius, and the outside brightness is 10; Scenario 2 corresponds to the situation where the goal remains the same after the context change but the performance indicators are changed, i.e., the outside moisture is changed to 0.3 and the temperature is changed to 20 degrees Celsius; Scenario 3 corresponds to the process of re-decision making after the goal changes, i.e., the user behavior changes to sleeping.

\subsubsection{Results}

The SASO properties utilizing our method and baseline method are detailed in Table \ref{tab:saso-sr}. In Scenarios 1 and 2, our approach delivers better results for \emph{temperature} and \emph{moisture}. This is because the softgoal \emph{good body comfort} has a higher priority in these contexts, leading the controller to allocate more resources to ensure comfort. However, in Scenario 3, following a change in context \emph{user activity}, the priority shifts to the softgoal \emph{low budget}, prompting the controller to opt for lower costs. Additionally, the reference value for \emph{brightness} changes from 100 to 0, our method can quickly adapts, achieving improved results. Nonetheless, it is imperative to recognize that MPC problems commonly encompass multiple outputs, each with distinct reference goals, necessitating a balanced trade-off in the overall objective function that may not achieve optimization across all performance indicators.

\begin{table}
  \caption{SASO property for SmartRoom case}
  \label{tab:saso-sr}
  \begin{tabular}{l|cc|cc|cc}
    \toprule
    \multirow{2}{*}{Metric}  & \multicolumn{2}{c}{Scenario 1} \vline & \multicolumn{2}{c}{Scenario 2} \vline & \multicolumn{2}{c}{Scenario 3}\\
    & ours & baseline & ours & baseline & ours & baseline \\
    \midrule
    Stability & True & True & True & True & True & True\\
    Accuracy (cost)& 26.375 & \textbf{15.125} & 26.687 & \textbf{16.062} & \textbf{11.875} & 26.688\\
    Accuracy (temperature)& \textbf{2.342} & 2.844 & \textbf{2.182} & 2.920 & 2.994 & \textbf{2.358} \\
    Accuracy (moisture)& \textbf{0.0212} & 0.0254 & \textbf{0.0154} & 0.0313 & 0.0357 & \textbf{0.0275} \\
    Accuracy (brightness)& 0 & 0 & 0 & 0 & \textbf{10} & 100\\
    Settling time & 1T & 1T & 2T & 2T & 1T & 1T\\
    Overshoot (cost)& 32 & 32 & 32 & 32 & \textbf{25} & 32\\
    Overshoot (temperature)& 3.636 & \textbf{3.574} & \textbf{2.431} & 3.928 & 4.285 & \textbf{3.636} \\
    Overshoot (moisture)&\textbf{0.0490} & 0.5005 & \textbf{0.0344} & 0.0606 & 0.0547 & \textbf{0.0447}\\
    Overshoot (brightness)& 0 & 0 & 0 & 0 & \textbf{10} & 100\\
  \bottomrule
\end{tabular}
\end{table}

\begin{figure*}[!t]
\centering
\subfloat[RQ1]{\includegraphics[width=0.48\textwidth]{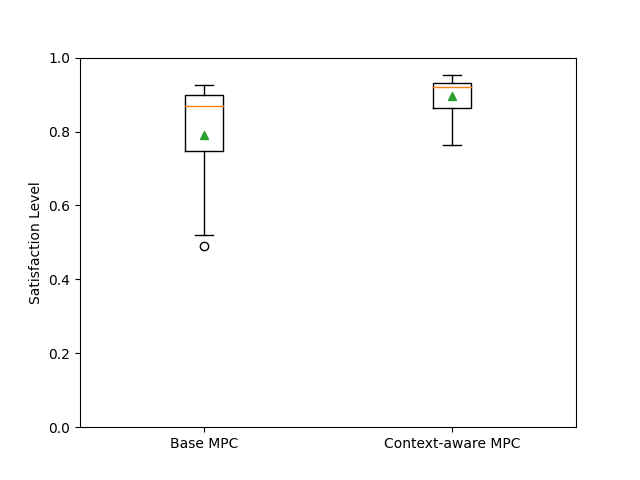}
\label{fig:srrq1}}
\hfil
\subfloat[RQ2]{\includegraphics[width=0.48\textwidth]{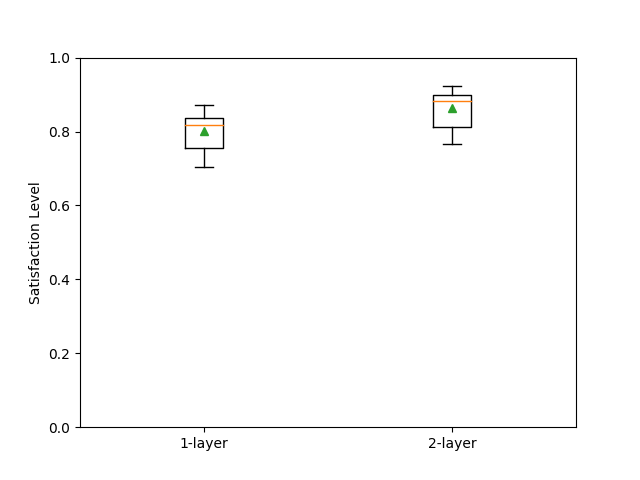}
\label{fig:srrq2}} 
\caption{Satisfaction Level of random environment cases in RQ1 and RQ2}
\label{fig:srres}
\end{figure*}

For RQ1, we randomly generated 20 distinct datasets of context variables with durations of 48 time steps. To eliminate the influence of the Meta Control Layer, we ensured that any context variables associated with it remained static during runtime. For example, the context variable \emph{user activity} remained static. The average satisfaction value was collected for each dataset, and the results are displayed in Figure \ref{fig:srrq1}, with triangular points denoting mean values. The base MPC’s mean satisfaction from the 20 sets was 0.78 (standard deviation of 0.16), while that of the context-aware MPC stood at 0.87 (standard deviation of 0.048). We conduct the Wilcoxon signed-rank test on the results, which is a non-parametric statistical hypothesis test for paired non-normally distributed data. The null hypothesis $H_0$ is that the two paired satisfaction data come from the same distribution. The one-sided alternative hypothesis $H_1$ is that they come from different distributions. The p-value of the Wilcoxon signed-rank test is 2.5e-10. Hence, we would reject the null hypothesis at a confidence level of 0.05, concluding that there is an improvement in the effectiveness of context-aware MPC.
Evidently, the base MPC failed to develop an accurate system model for the Smart Room case, due to overlooking the influence of context variables on performance indicators. This diminished the effectiveness of the adaptation. Furthermore, the context-aware MPC approach demonstrated a considerably reduced standard deviation relative to the Base MPC, maintaining satisfactory outcomes even in worst-case scenarios.

For RQ2, we additionally generated 20 different datasets of context variables, but without any constraints. The average satisfaction value of each dataset was recorded, and the results are presented in Figure \ref{fig:srrq2}. The single control layer scored a mean satisfaction of 0.80 (standard deviation of 0.049), while the two control layers had a mean satisfaction of 0.86 (standard deviation of 0.047). We also conducted the Wilcoxon signed-rank test on the satisfaction value, and the results denote a significant improvement as well. 
This indicates the Meta Control Layer's capability to fine-tune the goal model to the context, handling the effects of the context on goals, tasks, softgoal preferences, and softgoal quality functions, thus improving the adaptation effectiveness.

\begin{figure*}[!t]
\centering
\subfloat[User activity]{\includegraphics[width=0.45\textwidth]{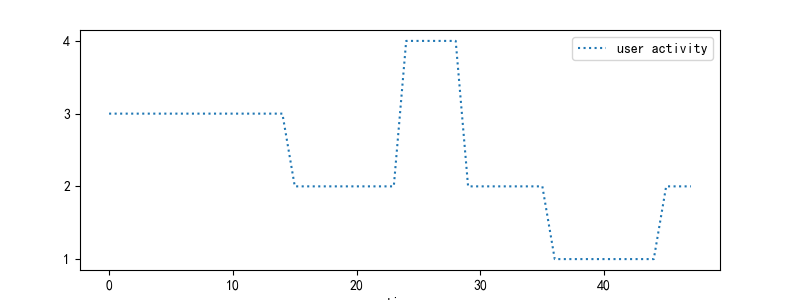}
\label{fig:exp1-3-1}} \\
\hfil
\subfloat[Goal]{\includegraphics[width=0.45\textwidth]{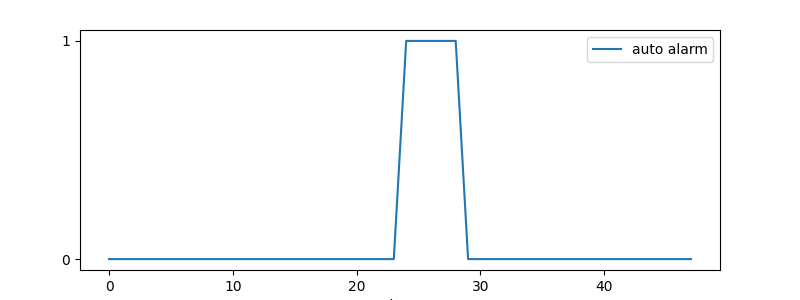}
\label{fig:exp1-3-2}} 
\hfil
\subfloat[Task]{\includegraphics[width=0.45\textwidth]{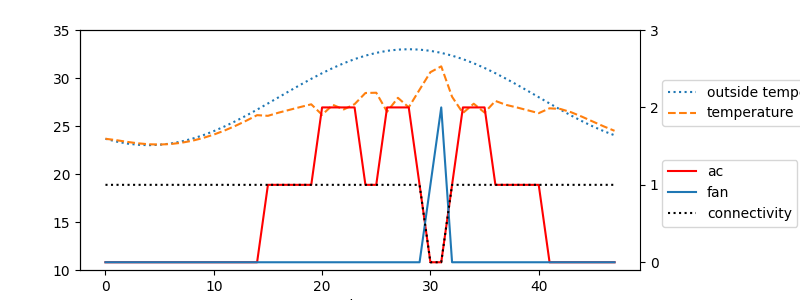}
\label{fig:exp1-3-3}} \\
\hfil
\subfloat[Softgoal weights]{\includegraphics[width=0.45\textwidth]{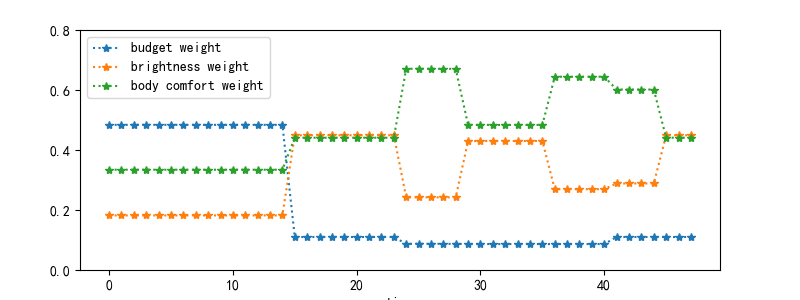}
\label{fig:exp1-3-4}} 
\hfil
\subfloat[Softgoal quality function]{\includegraphics[width=0.45\textwidth]{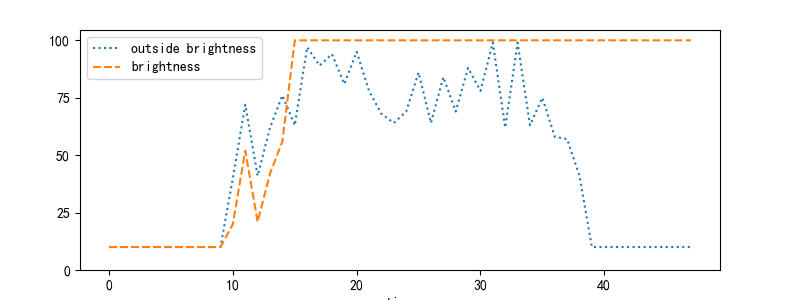}
\label{fig:exp1-3-5}} 
\caption{Results of different contextual influence}
\label{fig:exp1-3}
\end{figure*}

Figure \ref{fig:exp1-3} illustrates a typical scenario. Figure \ref{fig:exp1-3-1} outlines the variable, \emph{user activity}, over time. Here, values 1-4 represent user states: idle, work, sleep, and away, respectively.
Figures \ref{fig:exp1-3-2} to \ref{fig:exp1-3-5} show how varied contexts influence the system and the subsequent system adaptations. To elaborate: \emph{Goal Activation of Safety Alert.} Figure \ref{fig:exp1-3-2} highlights the activation of the safety alert goal at time slice 24 due to the context c4, adjusting the related control parameter to a value of 1. The baseline method without the Meta control layer may ignore the change in goal brought about by this context change and fail to achieve this new activated goal.
\emph{Task Selection of Temperature Control. } In Figure \ref{fig:exp1-3-3}, when the context c1 holds, the system employs the control parameter \emph{ac} to regulate temperature. However, as the narrative reaches time slice 30, the context c1 dissipates, prompting the system to switch to using the \emph{fan} control parameter to adjust the temperature. The baseline method does not respond to this change in time to adjust to other control parameters, resulting in higher temperatures.
\emph{Softgoal Weight Variation.} Figure \ref{fig:exp1-3-4} illustrates the dynamic nature of the softgoal weights. A notable inflection point occurs at time slice 24 when the context c5 is in effect. This indicates that the prevailing temperature is approaching an extreme, leading to a surge in the priority of the \emph{good body comfort} softgoal. The baseline method, on the other hand, maintains a uniform softgoal priority, which may reduce user satisfaction when the context changes.
\emph{Softgoal Quality Function Variation of Brightness Control.} Figure \ref{fig:exp1-3-5} details the management of the \emph{brightness} performance indicator. Under the influence of context c2, a subdued light intensity is preferred. In alternative scenarios, brighter lighting is favored. This distinction becomes evident at time slice 15, where the brightness's reference value transitions from 0 to 100. The baseline method that does not take this into account would still choose the original wrong reference value, leading to the opposite effect.

For RQ3, during the process of conducting the experiments for RQ2, we collected the t1 and t2 as described in the previous section. The average value of t1 is 0.01s and the average value of t2 is 0.38s, which is completely acceptable in the Smart Room case.

\subsection{SWIM}
\subsubsection{Experiment setting}
\begin{figure*}[!t]
\centering
\includegraphics[width=0.9\textwidth]{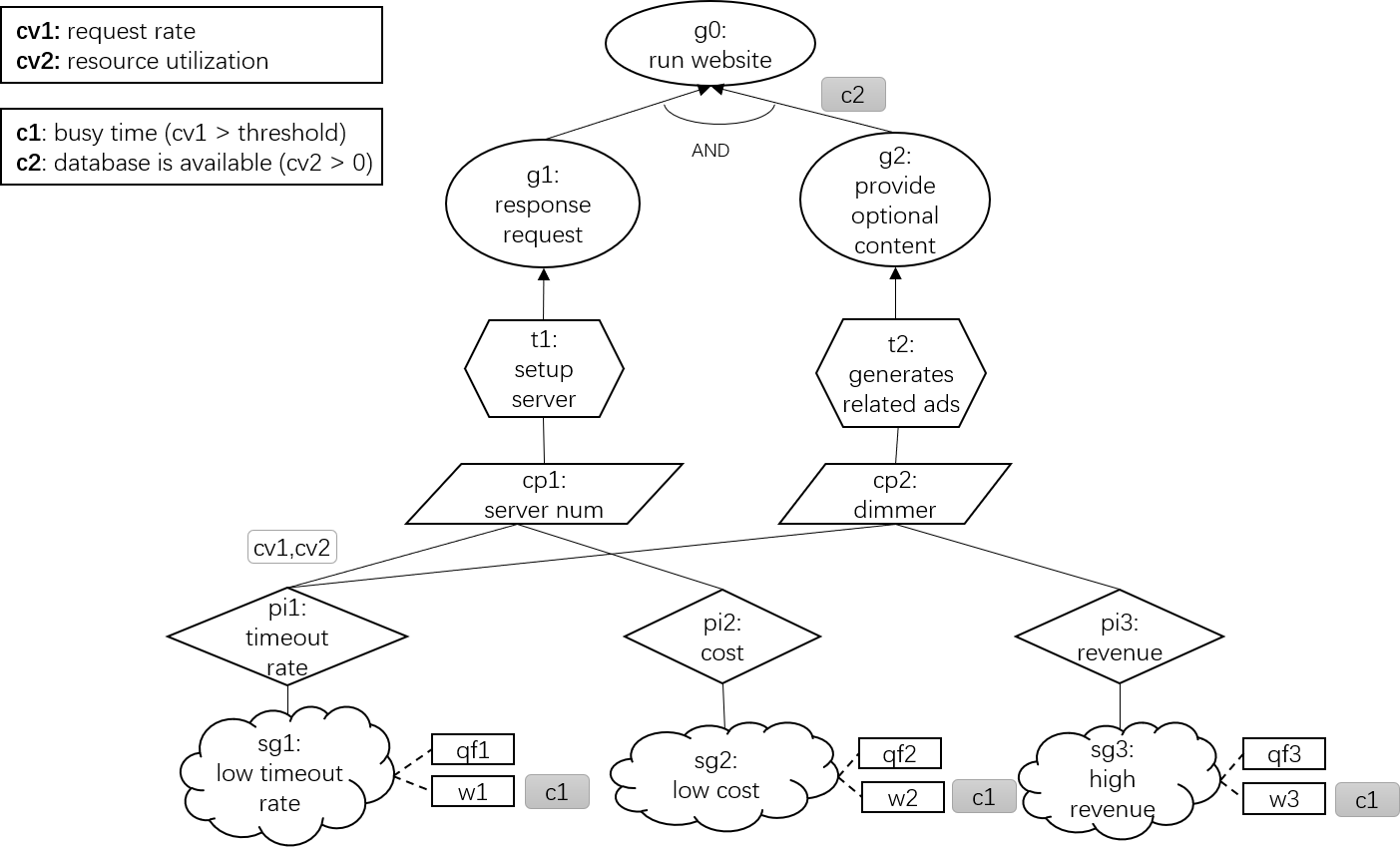}
\caption{Contextual Goal Model for SWIM.}
\label{fig:cgm-swim}
\end{figure*}

SWIM simulates a general multi-layer web application similar to Znn.com and RUBiS, such application consists of a web server layer and a database layer. The web server layer receives requests from clients. A load balancer is used to support multiple servers in the web tier. When a client requests a web page using a browser, the web server that handles the request accesses the database tier to get the data needed to render the web page. The contextual goal model of the SWIM case is shown in Figure \ref{fig:cgm-swim}.

The root goal of SWIM is to ensure the continuous operation of the site. We identify two sub-goals: responding to user requests and providing additional content (such as advertisements) for enhanced profitability. However, the three softgoals, specifically \emph{low timeout rate, low cost, and high revenue}, are profoundly influenced by context variables. The two crucial context variables, \emph{request rate} and \emph{resource utilization rate}, are time-variant. An increase in the request rate leads to a longer request queuing time, and a higher resource utilization rate results in an extended service time for a request. Both of these factors contribute to a significant rise in the timeout rate. Thus, predicting changes in these two context variables is valuable for estimating timeout rates and is the foundation for subsequent effective proactive adaptation. On the basis of the two context variables, we define two contexts that influence the goals of the software system, context c1 indicates a busy scenario, thus the priority of the softgoals \emph{low cost} and \emph{low timeout rate} decreases, while the priority of the softgoal \emph{high revenue} rises; context c2 indicates that the resources related to the recommendation of the content are unavailable, and therefore optional content can not be provided. 

SWIM has two ways to deal with context changes. First, the system can add or remove servers from the server pool connected to the load balancer. Second, the system can choose whether to provide optional content (such as advertisements or recommendations), more optional content will increase the computational overhead of the system. This means the SWIM system can cope with the increase in workload by degrading the user experience, namely, brownout-compliant \cite{klein2014brownout}. SWIM uses a "dimmer" value to control the proportion of optional content. 

We similarly start with three scenarios to verify the SASO properties. Scenario 1 corresponds to the adaptation from the initial state to stability, with an initial user request rate of 30/s, resource utilization of 0, an initial number of servers of 2, and a dimmer of 0.5; Scenario 2 involves changing the user request rate and resource utilization during operation, in this case to 60/s and 0.5; and Scenario 3 involves changing the resource availability, the \emph{provide optional content} goal no longer needs to be achieved.

To simulate a real scenario, we used ClarkNet \cite{arlitt1996web} and WorldCup’98 \cite{arlitt2000workload} access data as request sources, which are real data collected from network nodes and websites and have been used in many web-related studies. We take the disk usage data characteristics of a real website as a reference to generate the corresponding resource utilization rate data for these two datasets. 

\begin{figure*}[!t]
\centering
\subfloat[Low Timeout Rate]{\includegraphics[width=0.3\textwidth]{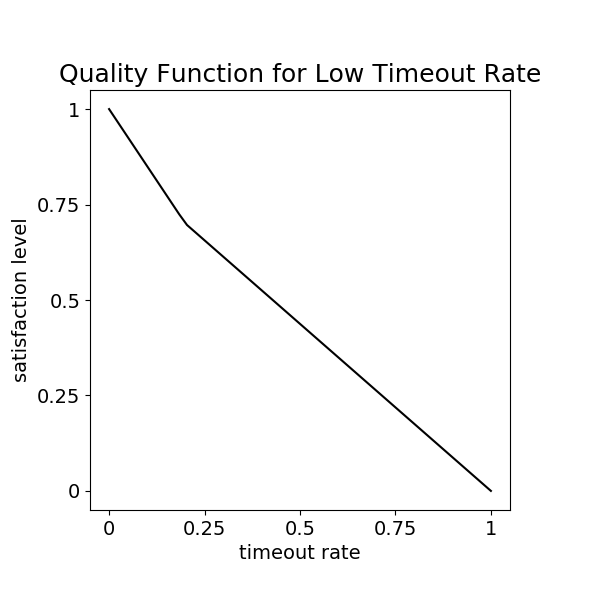}
\label{fig_swimqf1}}
\hfil
\subfloat[Low Cost]{\includegraphics[width=0.3\textwidth]{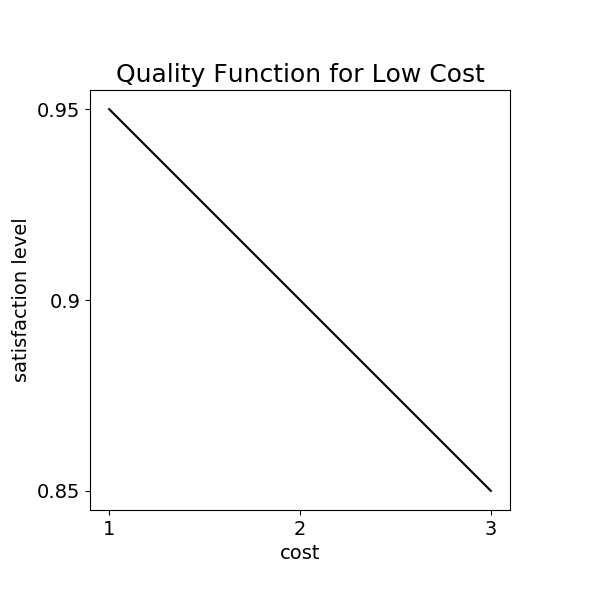}
\label{fig_swimqf2}} 
\hfil
\subfloat[High Revenue]{\includegraphics[width=0.3\textwidth]{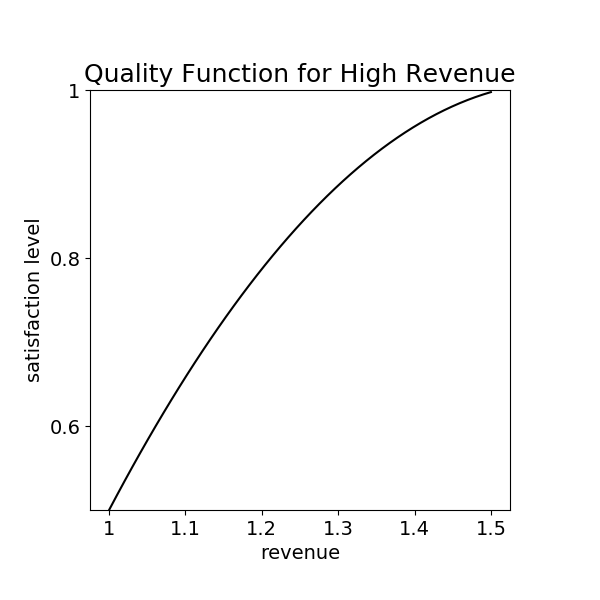}
\label{fig_swimqf3}} 
\caption{Quality Functions for Softgoals}
\label{fig_swimqf}
\end{figure*}

The default quality function for each softgoal is depicted in Figure \ref{fig_swimqf}, the value of the performance indicator \emph{timeout rate} can be obtained directly from the runtime data, while the values of the performance indicators \emph{cost} and \emph{revenue} are calculated based on the current value of control parameters \emph{server num} and \emph{dimmer}. We refer to the calculation of utility values in SWIM, which is $revenue = (1 - dimmer) * basicRevenue + dimmer * optRevenue, cost = 1 - serverCost * numServers$. The default weight for each softgoal is set equal. The satisfaction level of the system is calculated as described in the previous section. In our experiment, the threshold of the peak request rate is $MINREQ+0.6*(MAXREQ - MINREQ)$, where the $MINREQ, MAXREQ$ indicate the minimal request rate and maximum request rate in the trace, and the three softgoals priorities are adjusted to $w_1=0.2,w_2=0.2,w_3=0.6$ during the peak period; for each trace, we chose 10 time slices that are not in the peak period as the resource unavailability time.

\subsubsection{Results}
Table \ref{tab:saso-swim} shows the SASO properties in the three scenarios. In Scenario 1, both methods produce nearly identical results, this is due to the fact that the workload in this scenario is very small and the context prediction capability of our method is not effective in a stable environment. In Scenario 2, as the workload increases, our method prioritizes the timeout rate and revenue metrics due to the lower priority assigned to cost in this scenario. Therefore, our controller chose to increase the control parameters \emph{server num} and \emph{dimmer} to get a higher level of satisfaction. In Scenario 3, with the dimmer value limited to 0, our method effectively reduces the number of servers to achieve lower costs, demonstrating its ability to adapt when goals shift.

\begin{table}
  \caption{SASO property for SWIM}
  \label{tab:saso-swim}
  \begin{tabular}{l|cc|cc|cc}
    \toprule
   \multirow{2}{*}{Metric}  & \multicolumn{2}{c}{Scenario 1} \vline & \multicolumn{2}{c}{Scenario 2} \vline & \multicolumn{2}{c}{Scenario 3}\\
    & ours & baseline & ours & baseline & ours & baseline \\
    \midrule
    Stability & True & True & True & True & True & True\\
    Accuracy (timeout rate)& 3.13e-3 & 3.13e-3 & \textbf{1.05e-3} & 1.88e-3 & 0 & 0\\
    Accuracy (cost)& 2 & 2 & 3 & \textbf{2} & \textbf{1} & 2\\
    Accuracy (revenue)& \textbf{0.626} & 0.630 & \textbf{0.632} & 0.854 & 1 & 1\\
    Settling time & 3T & 3T & 2T & 2T & 1T & 1T\\
    Overshoot (timeout rate)& 0.157 & \textbf{0.156} & \textbf{0.016} & 0.058 & 0 & 0\\
    Overshoot (cost)& 2 & 2 & 3 & \textbf{2} & 2 & 2\\
    Overshoot (revenue) & \textbf{0.706} & 0.709 & \textbf{0.660} & 0.930 & 1 & 1\\
  \bottomrule
\end{tabular}
\end{table}

\begin{figure*}[!t]
\centering
\subfloat[WorldCup98]{
\includegraphics[width=0.48\textwidth]{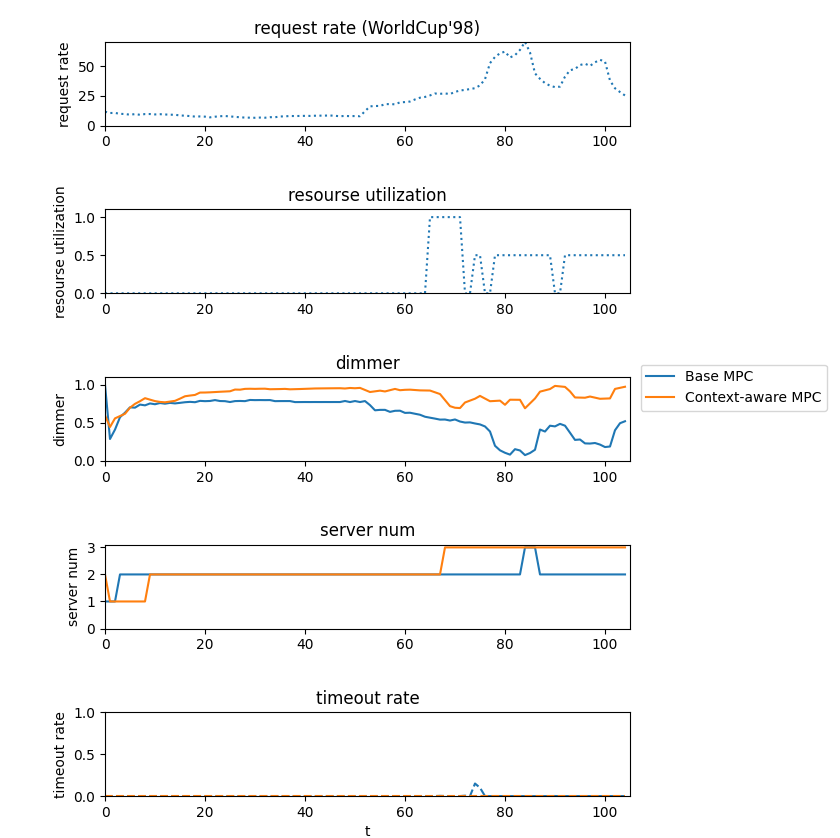}
}
\hfil
\subfloat[ClarkNet]{
\includegraphics[width=0.48\textwidth]{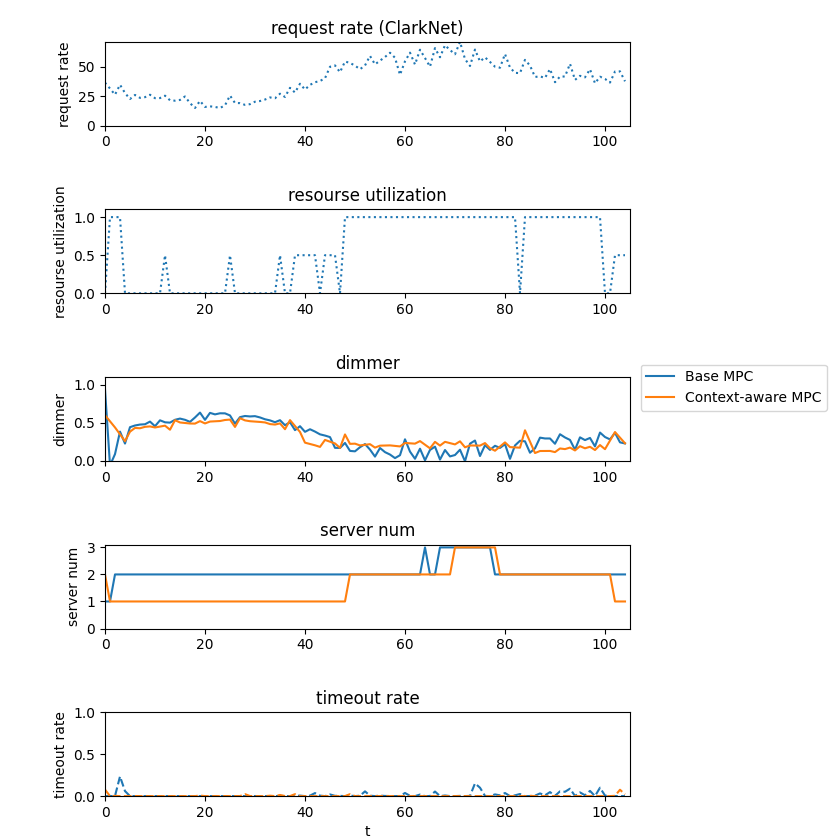}
} 
\caption{Running results of two traces in RQ1}
\label{fig_swimres1}

\end{figure*}

\begin{figure*}[!t]
\centering
\subfloat[WorldCup98]{
\includegraphics[width=0.48\textwidth]{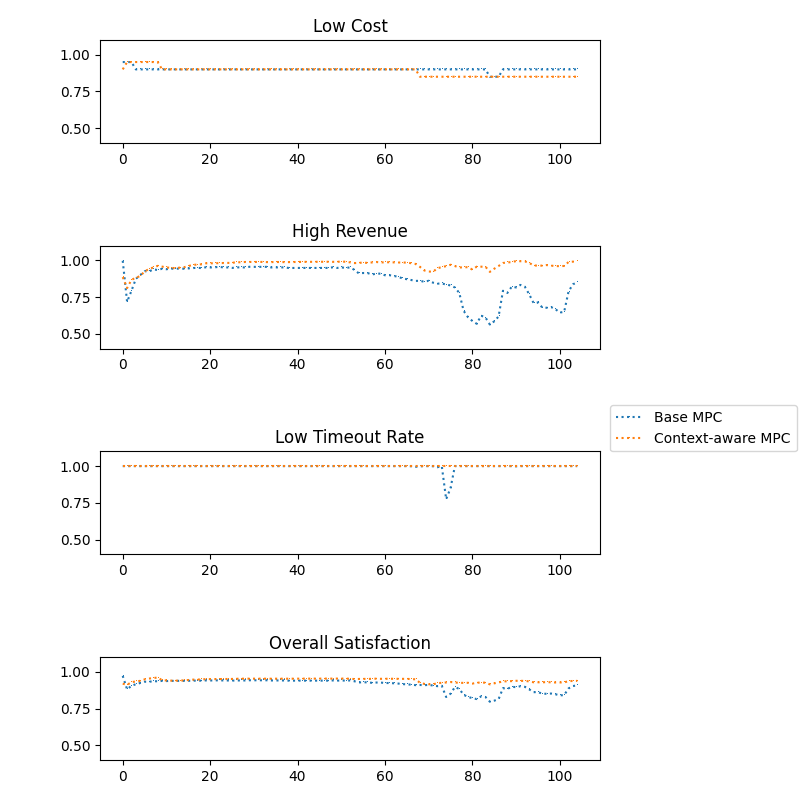}
}
\hfil
\subfloat[ClarkNet]{
\includegraphics[width=0.48\textwidth]{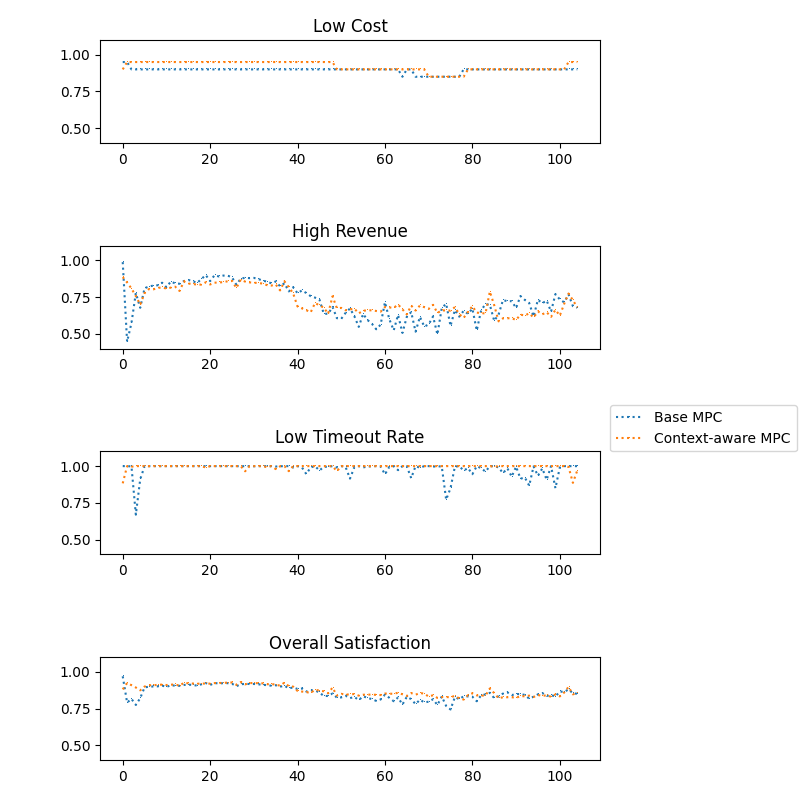}
} 
\caption{Satisfaction level of two traces in RQ1}
\label{fig_swimres2}

\end{figure*}

For RQ1, we compare the effectiveness between the base MPC and the context-aware MPC, and we consider the influence of the context variable requests rate and resource utilization on the SWIM performance indicators. We feed the SWIM with the two request traces and keep the contextual goal model fixed. In the SWIM case, it is difficult to predict the actual changes of the system based on the nominal system model, the base MPC method cannot achieve controllability. Thus we modified the base MPC method according to the method in \cite{moreno2017comparing}, adding contextual factors, but no prediction. The results are shown in Figure \ref{fig_swimres1},\ref{fig_swimres2}. We can find that in both scenarios, our method can achieve a better satisfaction level. In the WorldCup'98 case, the average satisfaction of the base MPC was 0.910 and the average satisfaction of the context-aware MPC was 0.942. In the Clarknet case, the average satisfaction of the base MPC was 0.857 and the average satisfaction of the context-aware MPC was 0.871. In the WorldCup’98 traces, due to the small load pressure in the early stage of the experiment, all methods can achieve good utility values. Around 50 time units into the experiment, the request rate suddenly begins to rise. Due to the proactive context prediction capability of our method, we anticipate this surge and preemptively increase the number of servers. This proactive adjustment prevents some timeouts before they can occur. In contrast, the base MPC method, which lacks any form of environmental prediction, reacts only after the request rate has already increased, leading to a rise in the timeout rate because the server resources are not scaled up in time.

\begin{figure*}[!t]
\centering
\subfloat[WorldCup98]{
\includegraphics[width=0.48\textwidth]{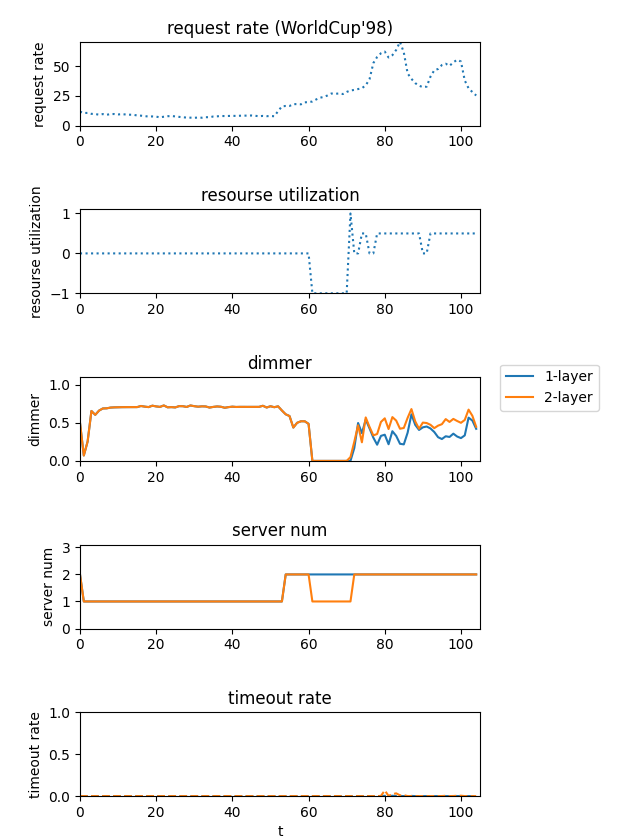}
}
\hfil
\subfloat[ClarkNet]{
\includegraphics[width=0.48\textwidth]{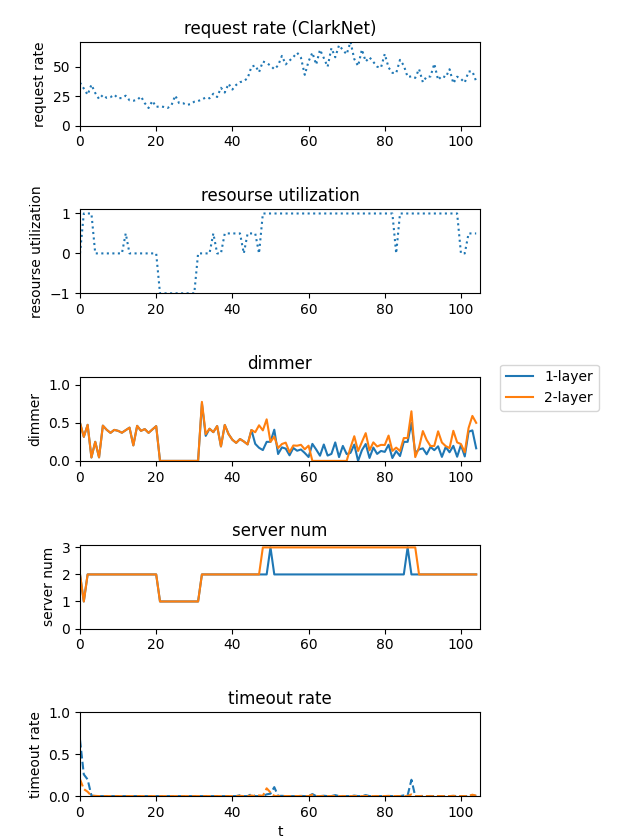}
} 
\caption{Running results of two traces in RQ2}
\label{fig_swimres3}

\end{figure*}

\begin{figure*}[!t]
\centering
\subfloat[WorldCup98]{
\includegraphics[width=0.48\textwidth]{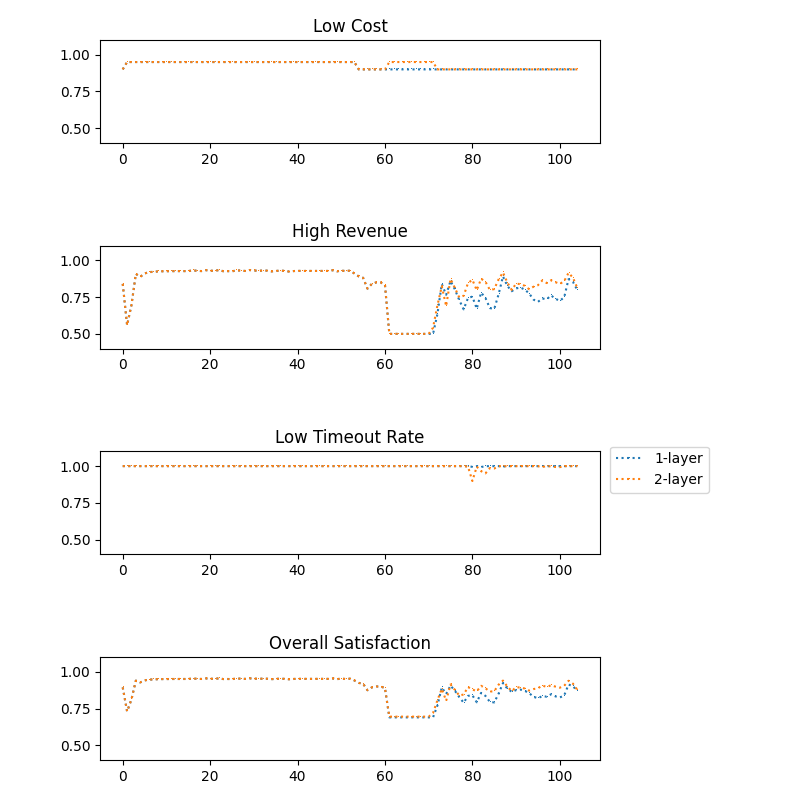}
}
\hfil
\subfloat[ClarkNet]{
\includegraphics[width=0.48\textwidth]{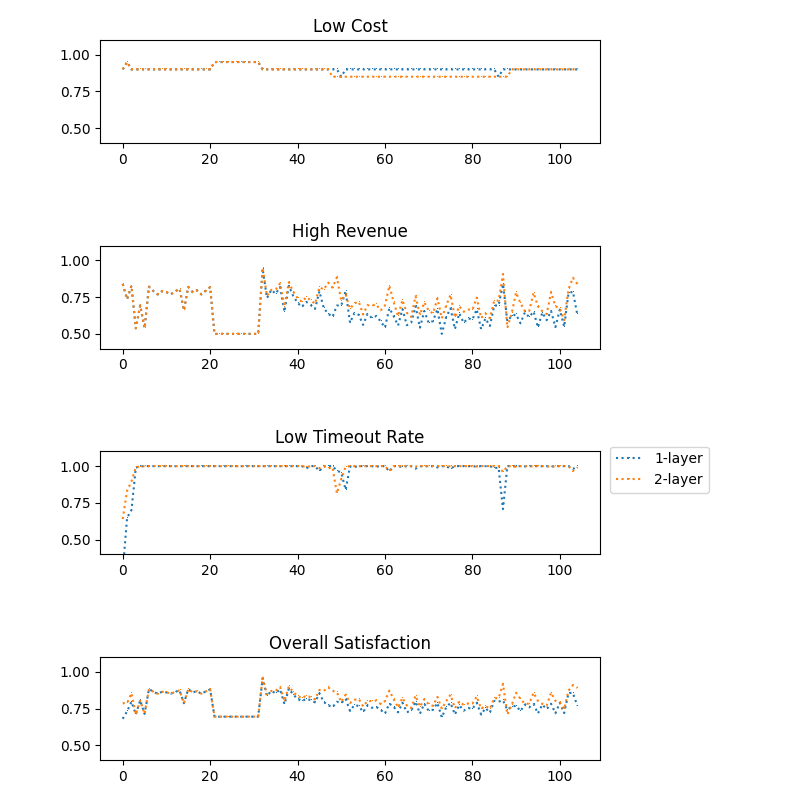}
} 
\caption{Satisfaction level of two traces in RQ2}
\label{fig_swimres4}

\end{figure*}

For RQ2, we compare the effect between the context-aware MPC with or without Meta Control Layer. In SWIM, the Meta Control Layer deals with two main context changes that influence requirements. One is to adjust the priority of softgoals during peak user requests, when the priorities of \emph{low timeout rate} and \emph{low cost} are lowered while the priority of \emph{high revenue} is raised; the other is when the goal g2 cannot be accomplished when resources related to the recommended content are unavailable, so the valid value of control parameter \emph{dimmer} could only be 0. The results are shown in Figure \ref{fig_swimres3},\ref{fig_swimres4}. From the results, we can see that during the peak request rate, the two-layer adaptation approach has higher dimmer values, or more servers, due to the prioritization of the softgoals, which may result in a higher cost or higher timeout rate, but at the same time, it results in higher revenue and improved overall satisfaction. When the resources related to recommendations are not available, the two-layer adaptation approach is able to update the CGM and deactivate the goal \emph{provide optional content}, making the constraint of the value of the control parameter \emph{dimmer} in the MPC optimizer 0, thus SWIM can choose fewer servers in order to reduce costs. In contrast, the one-layer approach, which does not consider this context, computes invalid dimmer values that cannot be applied to the system.

Since we chose two real-world access data for the experiment, which is less than the requirement of Wilcoxon signed-rank test for the number of data, and thus could not perform testing. However, we took the results of each time slice in each set of access data as separate data for hypothesis testing and also obtained significant results.

For RQ3, during the process of conducting the experiments for RQ2, we collected the t1 and t2 as described in the previous section. The average value of t1 is 0.13s and the average value of t2 is 0.04s, which is completely acceptable in the SWIM case.

\subsection{Discussion}

Through two experiments, we can show that our two-layer control approach is able to effectively cope with the influence of context on the software system, and improve satisfaction through a small time overhead. Using context-aware model predictive control, the systems are able to select appropriate control parameters for proactive self-adaptation, and adding a Meta Control Layer enables adaptation of the MPC controller by adjusting the contextual goal model according to the context. 
Admittedly, the two-layer approach introduces some additional overhead. At runtime, there is a small time overhead associated with context variable prediction and goal adjustment. However, this time consumption is relatively minor compared to the planning time overhead inherent in MPC. At design-time, the need to collect data and construct the necessary models incurs additional development costs. Fortunately, there exists available guidance (e.g. available time series model library for context variable prediction and system identification for system model) that can be utilized to alleviate this extra workload. Consequently, within the realm of complex software systems that experience frequent context changes, the enhancement in utility provided by the two-layer approach becomes more alluring and worthy of consideration.

Internal validity: Firstly, a significant threat emerges from the formulation of adaptive rules at the Meta Control Layer. As these rules are closely related to the user's requirements, we use pre-designed rules directly in our experiments. However, ensuring the completeness and correctness of these rules is challenging due to potential limitations in expert domain knowledge, and inappropriate rules may adversely affect the system’s performance. Although we have discussed methods to aid rule design in previous sections, this aspect requires further investigation. Secondly, the contextual conditions during the experiment also play a vital role in determining the method's effectiveness. In the SmartRoom scenario, we mitigated this threat by randomly generating the environment. Conversely, in the SWIM case, we utilized real access traces. Nevertheless, the scarcity of access traces remains an issue that needs to be addressed. Furthermore, the selection of contextual variable prediction models and system models constitutes another influential factor. In this paper, we opted for the basic models that are prevalently used in research. These models, however, are not customized or optimized for specific cases. It is reasonable to assume that more precise and case-specific models could potentially yield more favorable and accurate results, highlighting the need for a more meticulous consideration and evaluation of model selection in future studies to enhance internal validity.

External validity: To illustrate the generalizability of the approach, we chose two experimental cases, where the SWIM case is a simulation of a real website scenario and uses real access data, while the Smart Room case is a virtual scenario with a randomly generated runtime environment. Nonetheless, it is important to note that our approach incurs a certain amount of time overhead. This temporal cost may render it impractical for direct application in scenarios with stringent real-time demands. Furthermore, the context goal model utilized in our study is founded on control parameters and performance indicators. This foundation endows it with particular suitability for parametric adaptation. However, when it comes to system structure-based adaptation, an additional step of structure-to-parameter mapping becomes essential.

Conclusion validity: In order to eliminate threats to the process of drawing conclusions from experimental results, we use the Wilcoxon signed-rank test for significant difference between our method and baseline at 95\% confidence level to eliminate this threat. Nevertheless, it is essential to recognize that certain limitations exist in our approach. In particular, the utilization of softgoal satisfaction as a utility value in RQ1 and RQ2 presents some challenges. It is imperative that the satisfaction calculations we define closely align with user expectations. Ideally, a more comprehensive and accurate practice would involve the seamless incorporation of user feedback directly into the construction of the goal model.

\section{Related Work}
In the field of self-adaptive software systems, control-based approaches have had many applications. And MPC-based methods are often used for proactive self-adaptation. The CobRA method in \cite{angelopoulos2016model}\cite{angelopoulos2018engineering} combined MPC with requirement engineering. It expresses the goals and adjustable points of the software system through the extended requirement model, so as to control. The ProDSPL method in \cite{ayala2021prodspl} combined MPC and feature model to find the optimal configuration of key performance indicators through control theory while satisfying the constraints of the feature model. As mentioned above, the influence of the context on the system is rarely considered in these works, while our work uses a contextual goal model to describe the context influence and leverages a context-aware MPC method.

Concerning contextual requirements, the contextual goal model, as proposed in \cite{ali2010goal}, enriches the traditional goal model by incorporating contextual information. It introduces a method for contextual analysis, enabling the software system to discern and select the most fitting goal and goal decomposition in accordance with the prevailing context. 
In \cite{guimaraes2017planning} the concept of pragmatic goal is proposed, the influence of context on goal satisfaction is introduced, and the legitimate sequence of tasks is considered for planning. The literature \cite{alrajeh2020adapting} presents a method that automates the modification of requirement models for software systems in various context scenarios. This approach ensures the correctness, similarity, and localization of the requirement models, which can assist developers in automating the update of requirement models after contextual changes. 
Some work has also attempted to combine goal modeling with self-adaptive software systems.
For example, a goal-driven approach to taming uncertainty is proposed in \cite{felixsolano2019taminga}. This approach represents various types of uncertainty by adding annotations to the CGM. Combining the CGM and automatically generated symbolic formulae the system can synthesize adaptation policies and use these policies to guide runtime feedback control. In \cite{rodrigues2019goald}, a goal-driven deployment framework is proposed to support the automated deployment of computing resources to meet requirements in dynamic and heterogeneous computing environments. In the offline phase, GoalD prepares a deployment strategy by mapping goals to software components and environment dependencies. In the online phase, GoalD uses automated reasoning and scalable algorithms to select applicable components and create a deployment plan to achieve goals.
Our work advances beyond existing approaches by providing a comprehensive analysis of how context influences software system requirements. Unlike prior studies that primarily focus on how context changes affect goal activation and attainability\cite{alrajeh2020adapting,rodrigues2019goald}, our research extends these analyses to include dynamic shifts in softgoal priorities and variations in softgoal satisfaction. Additionally, our method systematically addresses a broader range of contextual factors impacting software systems. This allows for a more nuanced understanding of context-driven adaptations necessary for meeting diverse and evolving software requirements.

For multi-layer adaptation approaches, a classic three-layer model for self-adaptive systems was proposed in \cite{kramer2007selfmanaged}. It consists of three layers: component control, change management, and goal management, where the component control layer contains mechanisms to adjust the behavior of the system; the change management implements a MAPE loop to select the appropriate plan to adapt to the bottom layer, and the goal management also implements a MAPE loop to maintain the appropriate goal. MORPH, a reference architecture for behaviour and configuration self-adaptation, was presented in \cite{braberman2015morph}. It allows both independent reconfiguration and behaviour adaptation building on the extensive work developed but also allows coordinated configuration and behavioural adaptation to accommodate for complex self-adaptation scenarios. Based on MORPH, in \cite{caldas2020hybrid}, a two-layered adaptive manager was designed to satisfy the software requirements with parameters optimization through regression analysis and evolutionary meta-heuristic. In the adaptive manager, a strategy manager is used to synthesize adaptation strategies while a strategy enactor is used to enforce actions through control-theoretical principles. Both components are equipped with AI-based optimization techniques. In \cite{li2023preference} authors propose a two-layer self-adaptation framework. In the high-level layer, the system monitors for and dynamically updates the preference in response to a correction by the user. In the lower layer, the system will re-adjust the adaptive strategy to be executed by the automated planner based on the new preference. Our approach is similar to the three-layer model, using a Meta Control Layer (goal management layer in Kramer's model) to handle the influence of context on requirements.

\section{Conclusion}
In conclusion, we present an innovative two-layer adaptation framework grounded in a contextual goal model and context-aware MPC, addressing the contextual influence on software systems. By directly integrating the effect of context on goals, tasks, softgoals, and performance indicators, the contextual goal model offers a comprehensive and precise portrayal of software system requirements. The two-layer adaptation approach distinguishes adaptation challenges and separates adaptation concerns, providing a flexible and adaptive framework for dealing with dynamic contexts and enhancing the software system’s performance. 

In considering the potential limitations and opportunities for enhancement in our approach, several key aspects emerge. Initially, the precision of the system model is fundamental. Our current approach leans on a linear dynamic model of the software system. Although the integration of a Kalman filter aids in counteracting specific perturbations, the efficacy of this model is constrained, particularly in scenarios where the software system exhibits nonlinear characteristics. The potential for refining the model's accuracy and adaptability underscores a vital area for future advancement. 
Subsequently, there lies an expansive opportunity in the exploration of automated techniques for predicting user intentions and generating adaptation rules in Meta Control Layer. Potential advancements, perhaps leveraging machine learning algorithms informed by user feedback, or data analytics applied to system performance across diverse contexts, could offer enriched insights and enhance predictive accuracy. The evolution of these predictive mechanisms can significantly augment the completeness and correctness of the adaptation rules. 
Lastly, our current approach does not facilitate contextual prediction within the Meta Control Layer, rendering the requirements adaptation process reactive. This design decision is influenced by the risks associated with inaccuracies in requirement adaptation, prompted by erroneous predictions, and the consequential significant challenges that could ensue. Nevertheless, the prospect of integrating contextual prediction to evolve the requirement adaptation process from reactive to proactive is a compelling area for exploration. Developing strategies that harmonize the agility offered by proactive adaptations and the caution necessitated by the impacts of potential errors is central to advancing the robustness and efficacy of self-adaptive software systems.

\begin{acks}

\end{acks}

\bibliographystyle{ACM-Reference-Format}
\bibliography{mybib}

\end{document}